\documentclass[prc,twocolumn,twoside,floatfix]{revtex4}

\usepackage{graphicx,color,rotating}
\usepackage{amsmath,amssymb,bm}
\usepackage{times}

\newcommand{\eq}[1]{\begin{equation}#1\end{equation}}

\newcommand{\beq}{\begin{equation}}
\newcommand{\eeq}{\end{equation}}
\newcommand{\half}{\mbox{$\frac{1}{2}$}}

\newcommand{\six}{\mbox{$\frac{1}{6}$}}

\newcommand{\tfour}{\mbox{$\frac{1}{24}$}}

\newcommand{\ket}[1]{\ensuremath{\,|{#1}\rangle}}

\newcommand{\matrixe}[3]{\ensuremath{\langle{#1}|\,{#2}\,|{#3}\rangle}}
\newcommand{\expect}[1]{\ensuremath{\langle{#1}\rangle}}

\newcommand{\op}[1]{\ensuremath{#1}}

\renewcommand{\vec}[1]{\ensuremath{\bm{#1}}}

\newcommand{\HO}{\ensuremath{\op{H}}}
\newcommand{\TO}{\ensuremath{\op{T}}}
\newcommand{\VO}{\ensuremath{\op{V}}}

\newcommand{\POV}{\ensuremath{\vec{\op{P}}}}
\newcommand{\XOV}{\ensuremath{\vec{\op{X}}}}

\newcommand{\UCOM}{\ensuremath{\textrm{UCOM}}}
\newcommand{\intr}{\ensuremath{\textrm{int}}}
\newcommand{\refe}{\ensuremath{\textrm{ref}}}
\newcommand{\cm}{\ensuremath{\textrm{cm}}}
\newcommand{\elem}[2]{\ensuremath{{}^{#2}\text{#1}}}

\newcommand{\fm}{\ensuremath{\,\text{fm}}}
\newcommand{\MeV}{\ensuremath{\,\text{MeV}}}

\newcommand{\symboldiamond}[1][black]{{\color{#1}$\blacklozenge$}}

\newcommand{\symbolcircle}[1][black]{{\color{#1}\large$\bullet$}}

\newcommand{\symboldiamondopen}[1][black]{{\color{#1}$\lozenge$}}

\newcommand{\symbolcircleopen}[1][black]{{\color{#1}\large$\circ$}}

\definecolor{FGViolet}{rgb}{0.61,0.32,0.61}
\definecolor{FGDarkBlue}{rgb}{0,0,0.6}
\definecolor{FGBlue}{rgb}{0,0,0.8}
\definecolor{FGLightBlue}{rgb}{0.2, 0.6, 0.8}
\definecolor{FGGreen}{rgb}{0.2,0.7,0.2}
\definecolor{FGLightGreen}{rgb}{0.4,1,0.4}
\definecolor{FGYellow}{rgb}{1,0.95,0}
\definecolor{FGOrange}{rgb}{0.95,0.5,0.1}
\definecolor{FGRed}{rgb}{0.8,0,0}
\definecolor{FGWhite}{rgb}{1,1,1}
\definecolor{FGLightGray}{rgb}{0.8,0.8,0.8}
\definecolor{FGGray}{rgb}{0.5,0.5,0.5}
\definecolor{FGDarkGray}{rgb}{0.3,0.3,0.3}
\definecolor{FGBlack}{rgb}{0,0,0}

\begin{document}

\title{\emph{Ab initio} coupled-cluster and configuration interaction calculations for $^{16}$O using
$\VO_{\UCOM}$}

\author{Robert Roth}
\email{robert.roth@physik.tu-darmstadt.de}

\affiliation{Institut f\"ur Kernphysik, Technische Universit\"at Darmstadt,
64289 Darmstadt, Germany}

\author{Jeffrey R. Gour}

\author{Piotr Piecuch}
\email{piecuch@chemistry.msu.edu}

\affiliation{Department of Chemistry, Michigan State University, East Lansing, 
MI 48824-1322, USA}

\date{May 15, 2009}

\begin{abstract}   
Using the ground-state energy of \elem{O}{16} obtained with the realistic $V_{\UCOM}$ interaction as a test case, we present a comprehensive comparison of different configuration interaction (CI) and coupled-cluster (CC) methods, analyzing the intrinsic advantages and limitations of each of the approaches. In particular, we use the importance-truncated (IT) CI and no-core shell model (NCSM) schemes with up to 4-particle-4-hole (4p4h) excitations, with and without the Davidson extensivity corrections, as well as the size extensive CC methods with a complete treatment of one- and two-body clusters (CCSD) and a non-iterative treatment of connected three-body clusters via the completely renormalized correction to the CCSD energy defining the CR-CC(2,3) approach, which are all capable of handling larger systems with dozens of explicitly correlated fermions. We discuss the impact of the center-of-mass contaminations, the choice of the single-particle basis, and size-extensivity on the resulting energies. When the IT-CI and IT-NCSM methods include the 4p4h excitations and when the CC calculations include the 1p1h, 2p2h, and 3p3h clusters, as in the CR-CC(2,3) approach, we observe an excellent agreement among the different methodologies, particularly when the Davidson extensivity corrections are added to the IT-CI energies and the effects of the connected three-body clusters are accounted for in the CC calculations. This shows that despite their individual limitations, the IT-CI, IT-NCSM, and CC methods can provide precise and consistent \emph{ab initio} nuclear structure predictions. Furthermore, the IT-CI, IT-NCSM, and CC ground-state energy values obtained for \elem{O}{16} are in reasonable agreement with the experimental value, providing further evidence that the $V_{\UCOM}$ two-body interaction may allow for a good description of binding energies for heavier nuclei and that all of the methods used in this study account for most of the relevant particle correlation effects.
\end{abstract}

\pacs{21.60.Cs, 21.60.De, 31.15.bw, 27.20.+n }

\maketitle

\section{Introduction}

Recent developments in nuclear structure theory have lead to new perspectives for
a QCD-based \emph{ab initio} description of p-shell nuclei and beyond.
Significant advances have been made regarding the two main ingredients for an
\emph{ab initio} treatment of nuclear structure: the realistic interaction
defining the relevant Hamiltonian and the solution of the quantum many-body problem.
In addition to traditional realistic interactions, including the Argonne V18
\cite{WiSt95} or the CD Bonn potential \cite{Mach01}, the framework of chiral
effective field theory \cite{eft1,eft2}
has been used to construct consistent two- and many-nucleon
interactions based on the relevant degrees of freedom and the symmetries of QCD \cite{entem2002,EnMa03,EpNo02}.

Based on these bare interactions, several methods have been developed to adapt
the nuclear Hamiltonian to the limited model spaces available in practical many-body
calculations. Different conceptual frameworks ranging from the $V_{\text{low}k}$
renormalization group method \cite{BoKu03} and the similarity renormalization group
scheme \cite{BoFu07,HeRo07,RoRe08}
to the unitary
correlation operator method \cite{FeNe98,NeFe03,RoNe04,RoHe05} are being used to
construct soft, phase-shift equivalent interactions, which exhibit superior
convergence properties when employed in many-body calculations.

Bare or transformed interactions are the starting point for modern computational approaches for
the \emph{ab initio} solution of the nuclear many-body problem. Two methods have become particularly successful
in this area, namely, the Green's function Monte Carlo approach \cite{PiWi01,PiWiPRL,PiVa02,PiWi04,Pi2005}
and the no-core shell model \cite{NaVa00,ncsm2000a,ncsm2000b,ncsm2001,ncsm2002a,ncsm2002b,ncsm2002c,ncsm2003,%
ncsm2004a,ncsm_HF,ncsm2004b,ncsm2004c,ncsm2005a,ncsm2005b,ncsm2006a,ncsm2006b,ncsm2006c,NaGu07,RoNa07,ncsm2008,%
CaMa05}.
For nuclei up to $A\approx12$, detailed studies of ground and low-lying excited states,
including detailed spectroscopic information, were conducted with both methods using different
two-nucleon interactions and phenomenological three-nucleon forces.
Very recently, the first \emph{ab initio} no-core shell model calculations using
the chiral two- plus three-nucleon interactions have shown the great potential that this new
category of interactions may offer \cite{NaGu07}.

Although the Green's function Monte Carlo and no-core shell model
approaches are successful in describing nuclei up to the mid-p-shell region,
very few methods can provide a computationally tractable \emph{ab initio} description of
nuclei such as \elem{O}{16} or beyond. Among those are
the importance-truncated no-core shell model and the related
importance-truncated configuration interaction methods, and the coupled-cluster approach, which
have recently been used for nuclear structure calculations in the $A=16$ and $A=40$ mass regions
\cite{RoNa07,dean04,nucphysA_2005,bogdan1,bogdan2,bogdan4,kowalski04,o16prl,epja_2005,jpg_2005,prc2006,HaDe07}.
Extension of these methods toward heavier nuclei is presently in progress, within both the
\emph{ab initio} description and the traditional effective Hamiltonian approach (cf. Refs.
\cite{ni56_2007,ni55-57}).

The coupled-cluster and configuration interaction methodologies
that have been used in these recent developments
are complementary regarding their respective advantages and limitations.
This work aims at a direct and quantitative comparison of the results of large scale
\emph{ab initio} importance-truncated configuration interaction, importance-truncated no-core shell
model, and coupled-cluster calculations based on the same realistic Hamiltonian
using the example of the \elem{O}{16} ground state.
We focus on the most practical coupled-cluster and configuration interaction approaches
that have been specifically designed to handle larger systems with dozens of correlated particles and
larger single-particle basis sets.
The most essential formal differences, advantages, and limitations of the
importance-truncated configuration interaction, importance-truncated no-core shell model, 
and coupled-cluster methods are addressed and
their relevance for practical nuclear structure calculations is discussed. 
The nucleon-nucleon interaction used in this work is the $\VO_{\UCOM}$ potential
derived in the Unitary Correlation Operator Method (UCOM) \cite{FeNe98,NeFe03,RoNe04,RoHe05}. 

Following a description of the importance-truncated configuration interaction
and coupled-cluster methods in Secs. \ref{sec:ci} and \ref{sec:cc},
we discuss the benchmark results for \elem{O}{16} in Sec. \ref{sec:benchmark}. We
assess in detail the impact of center-of-mass contaminations and the role of the
single-particle basis, and compare the results of large-scale calculations based
on the importance-truncated configuration interaction, importance-truncated no-core shell model, 
and coupled-cluster methodologies with one another and with experiment.

\section{Importance-Truncated Configuration Interaction}
\label{sec:ci}

As a first class of methods, we consider different configuration interaction (CI) approaches.
The common element of all CI methods, which are frequently used in nuclear structure theory and quantum
chemistry, is that the eigenvalue problem for the Hamiltonian
is solved in a many-body basis of Slater determinants or symmetry-adapted configuration state
functions constructed from a given single-particle basis. In particular,
the so-called full CI approach of quantum chemistry employs a model space spanned by all
possible Slater determinants built from a finite set of single-particle orbits, providing the exact
solution of the Schr{\" o}dinger equation in that single-particle basis.
Unfortunately, the dimension of the full CI model space grows
factorially with the numbers of particles and single-particle orbits,
limiting full CI calculations to small systems and relatively small single-particle basis sets.

Truncated CI models offer a simple and natural way of reducing the dimension of the model space
and the prohibitive costs of full CI calculations.
In the simplest scheme that defines the single-reference truncated CI models,
only the excited determinants up to and including $m$-particle--$m$-hole ($m$p$m$h)
excitations from the reference determinant $|\Phi_{0}\rangle$, where $m$ is typically
much smaller than the number of active particles, are included in the Hamiltonian
diagonalization. By introducing the $k$p$k$h excitation
operators $C_{k}$, the ground state emerges then as 
\eq{
  \ket{\Psi_0} = \sum_{k=0}^{m} C_{k} \ket{\Phi_0} \;,
\label{truncated-ci}
}
where $C_{0} \ket{\Phi_0}$ is the reference determinant contribution to $\ket{\Psi_0}$ and
the $C_{k} \ket{\Phi_0}$ terms with $k=1,\ldots,m$ are the $k$p$k$h components of the wave function
included in the calculations. The popular approaches in this category of methods include
CISD (CI singles and doubles), CISDT (CI singles, doubles, and triples), and CISDTQ
(CI singles, doubles, triples, and quadruples), if
we use the quantum-chemistry acronyms, or CI($m$p$m$h) with $m=2$, 3, and 4, respectively, if we use the
particle-hole language of many-body physics.
Since these truncated CI methods lack size extensivity (i.e. unlinked diagrams are present in the
truncated CI wave function expansions, resulting in a potential loss of accuracy as the system
becomes larger), the Davidson extensivity corrections
\cite{davidson} are usually added to the resulting energies to alleviate the problem.

Another popular scheme to define truncated CI models, which is frequently used in quantum chemistry
and which is related to the importance-truncated nuclear CI methods considered in this work,
is to consider the 1p1h and 2p2h
excitations from the multi-configurational reference space spanned by a number of
determinants that provide a reasonable zero-order description of the many-body state or states
of interest, as in the multi-reference CI methods \cite{mrci,mrdci}. The multi-reference CI
approaches are particularly useful when a single reference determinant is a poor approximation
to the eigenstate(s) of interest and when one wants to accelerate convergence toward the
results of the exact diagonalization
by selecting the dominant contributions to the wave function due to 3p3h, 4p4h, and other
higher-order excitations. With the judicious choice of the multi-determinantal reference state,
the multi-reference CI methods can be very accurate, but
they often lead to long wave function
expansions. Thus, to further reduce
computer costs of multi-reference CI calculations,
one often considers reduced sets or subsets of the resulting many-body basis
states, which are usually defined through the internal contractions of configuration
state functions or the multi-configurational perturbation theory analysis coupled
with numerical thresholds to reject unimportant configurations.
The importance-truncation scheme used in this work uses similar ideas to those exploited
in the multi-reference CI approaches, while enabling one to incorporate the higher-order
$m$p$m$h excitations from the multi-determinantal reference space in a systematic fashion.
Again, since the multi-reference CI methods lack size extensivity,
the multi-reference generalizations of the Davidson extensivity correction
\cite{mrdci,sqdc1,sqdc2,sqdc3,sqdc4} are usually added to the resulting energies
(for the alternative, more intrinsic ways of approximately restoring size extensivity
in multi-reference CI calculations, see, e.g., Ref. \cite{szalay} and references therein).

In nuclear structure theory, the CI concept is widely used in the form of the diagonalization
shell model and all of its variants (see Ref. \cite{CaMa05} and references therein). We only
consider no-core calculations in this work, i.e. all particles are active. A full CI calculation
would use a complete model space spanned by all Slater determinants
that one can obtain for a given finite set of single-particle states,
e.g. the harmonic-oscillator single-particle states
up to a maximum principal quantum number $e_{\max}$, where $e = 2n+l$, with $n$ and $l$
representing the radial and angular momentum quantum numbers, respectively.
Other finite sets of single-particle states, e.g. those extracted from a
self-consistent Hartree-Fock calculation, can be employed in the CI calculations as well.
The nuclear truncated CI($m$p$m$h) approaches, where $m < A$, emerge then by considering
the suitable subsets of Slater determinants from the full CI model space,
including the reference determinant $|\Phi_{0}\rangle$ and all
$k$p$k$h excitations from $|\Phi_{0}\rangle$ with $k=1,\ldots,m$, as described above.
By systematically increasing $m$, one approaches the full CI limit in which
$m=A$, and by considering the $e_{\max} \rightarrow \infty$ limit, one
approaches the exact solution of the Schr{\" o}dinger equation for a given Hamiltonian.
In the following, we will reserve the term CI exclusively for the diagonalization shell model calculations
based on this specific form of the single-particle truncation.
 
A modification of the above full and truncated CI methods, which has been used in several highly successful
{\it ab initio} investigations of nuclei in recent years, is the no-core shell model (NCSM)
\cite{NaVa00,ncsm2000a,ncsm2000b,ncsm2001,ncsm2002a,ncsm2002b,ncsm2002c,ncsm2003,%
ncsm2004a,ncsm_HF,ncsm2004b,ncsm2004c,ncsm2005a,ncsm2005b,ncsm2006a,ncsm2006b,ncsm2006c,NaGu07,RoNa07,ncsm2008,%
CaMa05}.
The NCSM formalism uses a different definition of the complete model space than that employed in full CI.
Instead of considering all Slater determinants that one can obtain for a given set of the
harmonic-oscillator single-particle states up to a maximum principal quantum number $e_{\max}$,
in the NCSM approach
one includes those Slater determinants with an unperturbed excitation energy that does not
exceed $N_{\max}\hbar\Omega$, where $\Omega$ is the frequency associated with the
harmonic oscillator basis
\cite{NaVa00,ncsm2000a,ncsm2000b,ncsm2001,ncsm2002a,ncsm2002b,ncsm2002c,ncsm2003,%
ncsm2004a,ncsm_HF,ncsm2004b,ncsm2004c,ncsm2005a,ncsm2005b,ncsm2006a,ncsm2006b,ncsm2006c,NaGu07,RoNa07,ncsm2008,%
CaMa05}.
The exact solution of the Schr{\" o}dinger equation for a given Hamiltonian is systematically approached
by considering the $N_{\max} \rightarrow \infty$ limit.
Although the NCSM calculations rely on the
general CI concept of the diagonalization of the Hamiltonian used in all shell-model calculations,
in this work we distinguish between methods that employ all or some Slater determinants
resulting from the truncation of the single-particle space, which we continue calling the CI methods, and the
NCSM approaches that represent the special type of shell-model calculations which uses
the $N_{\max}\hbar\Omega$ model spaces. Typically, for the reasons explained below,
the NCSM calculations are performed in the
harmonic-oscillator bases (cf., however, Ref. \cite{ncsm_HF} for the Hartree-Fock based NCSM study).

For the application to nuclei, the NCSM model spaces offer a few important advantages
over the model spaces used in CI calculations. Unlike in quantum chemistry, where the issue of center-of-mass
does not exist due to the Born-Oppenheimer approximation, the proper description of self-bound
nuclei requires that special care is taken
regarding the center-of-mass contamination of intrinsic states.
Only the NCSM model spaces, in which the cutoff for the Slater determinants included in the calculations
is defined by setting up the maximum unperturbed excitation energy at $N_{\max}\hbar\Omega$,
allow for an exact separation of intrinsic and center-of-mass motions,
which is necessary to obtain translationally invariant intrinsic states, as long as a harmonic-oscillator single-particle basis is used. The CI truncations,
in which the cutoff for the Slater determinants included in the calculations is
typically defined by the number of major harmonic oscillator shells
corresponding to the maximum principal quantum number $e_{\max}$,
as defined above,
violate the separation of intrinsic and center-of-mass motions, resulting in the
center-of-mass contaminated states. The degree of center-of-mass contamination goes down
as $e_{\max} \rightarrow \infty$ and $m \rightarrow A$, but one cannot eliminate it mathematically, as in the NCSM case.
Furthermore,
the truncation with respect to the unperturbed excitation energy used in the NCSM calculations
automatically identifies the
most relevant Slater determinants in the wave function expansion,
as is clear from elementary perturbative arguments.
One of the questions addressed in this work is to examine the potential impact of
the center-of-mass contamination in truncated CI calculations on the results for ${}^{16}$O,
particularly when compared to other sources of errors that all
approximate many-body methods using computationally
tractable truncated model spaces carry.

For practical applications in nuclear structure calculations,
the dimensions of the NCSM and CI model spaces become prohibitively large,
even in relatively small systems, including the ${}^{16}$O nucleus examined in this work,
and even when we use the truncated CI
approaches. For example, in the aforementioned ${\rm CISDTQ \equiv CI(4p4h)}$ method
one has to deal with huge numbers of $\sim n_{o}^{3} n_{u}^{3}$ and
$\sim n_{o}^{4} n_{u}^{4}$ determinants of the 3p3h and
4p4h types and the iterative diagonalization steps that scale as $n_{o}^{4} n_{u}^{6}$, where
$n_{o}$ and $n_{u}$ are the numbers
of occupied and unoccupied single-particle states, respectively.
Therefore, in this study
we introduce an additional physically motivated truncation of the CI and NCSM model spaces,
which defines the importance-truncated (IT) CI and NCSM methods \cite{RoNa07,Roth08b}.
The IT-CI and IT-NCSM schemes are similar in the overall philosophy to the configuration
selection techniques used in the quantum-chemical MRD-CI model of Peyerimhoff and Buenker
\cite{mrdci,mrdci1,mrdci2}. Using the multi-configurational perturbation theory of Ref. \cite{SuRo04},
we estimate the importance of individual basis states for the description of a certain target state.
The amplitude with which a many-body basis state $\ket{\Phi_{\nu}}$ contributes in the first-order
perturbation theory to an initial approximation $\ket{\Psi_\refe}$ of the target state of
interest, e.g. the ground state $\ket{\Psi_{0}}$, defines an \emph{a priori} importance measure
\eq{
  \kappa_{\nu} 
  = -\frac{ \matrixe{\Phi_\nu}{\HO_{\intr}}{\Psi_\refe} }{ \epsilon_\nu - \epsilon_\refe } \;,
}
where $\epsilon_\nu-\epsilon_\refe$ is the unperturbed excitation energy resulting from
a M\o{}ller-Plesset-type partitioning of the Hamiltonian \cite{SuRo04}. With the intrinsic Hamiltonian
$\HO_{\intr}=\TO-\TO_{\cm}+V_{\UCOM}$ being a two-body operator, the non-zero
importance measures can be obtained only for the determinants $\ket{\Phi_{\nu}}$ that differ from
the reference state $\ket{\Psi_\refe}$ by a 2p2h excitation at most. 

In order to access higher-order $m$p$m$h excitations with $m > 2$,
we embed the concept of the importance measure into an iterative construction of the
importance-truncated model space. Starting from the lowest-energy Slater
determinant $\ket{\Phi_{0}}$ as an initial reference $\ket{\Psi_\refe}$,
all basis states with the importance measure $|\kappa_{\nu}| \geq \kappa_{\min}$,
where $\kappa_{\min}$ defines the importance threshold, are included in the model
space and the eigenvalue problem is solved. The resulting eigenvector, providing
an improved approximation for the target state, is used to define the reference
state $\ket{\Psi_\refe}$, which is now a multi-determinantal state,
for the second iteration and the above procedure is repeated.
In the limit $\kappa_{\min}\to0$, this iterative scheme converges to the full CI or NCSM space.
The perturbative character of the importance measure entails an $m$p$m$h hierarchy.
In the first iteration, only the determinants up to the 2p2h level are generated,
resulting in the IT-CI(2p2h) approximation.
In the limit $\kappa_{\min}\to0$, the IT-CI(2p2h) calculation recovers the full
${\rm CISD \equiv CI(2p2h)}$ model space. In the second iteration, up to 4p4h determinants are present,
leading to the ${\rm CISDTQ \equiv CI(4p4h)}$ model space in the $\kappa_{\min}\to0$ limit.
In this study, we restrict ourselves to two iterations of IT-CI and IT-NCSM, i.e. model
spaces up to the 4p4h level,
so that the resulting IT-CI(4p4h) wave function expansions are approximations to the CISDTQ wave functions.
We should emphasize that for \elem{O}{16} and for larger single-particle
basis sets used in this study, the full ${\rm CISDTQ \equiv CI(4p4h)}$ calculations would not be
possible due to the prohibitively large dimensions of the corresponding model spaces. The IT methodology
is among very few methods that enable one to incorporate 4p4h excitations in the CI calculations
for nuclei with larger numbers of correlated particles,
such as \elem{O}{16}, and large single-particle basis sets with hundreds of orbitals
in a computationally tractable fashion.

Eventually, as in the MRD-CI approach of Refs. \cite{mrdci,mrdci1,mrdci2},
we construct the importance-truncated model spaces and solve the corresponding eigenvalue problems
for a sequence of importance thresholds $\kappa_{\min}$ and extrapolate to $\kappa_{\min}\to0$.
All IT-NCSM and IT-CI results presented in Sec. \ref{sec:benchmark} are based on sequences of
calculations with the importance thresholds in the range $\kappa_{\min}=2\times 10^{-5}$ to
$15\times 10^{-5}$. In order to warrant a robust extrapolation result, we use additional
information from a second-order perturbative estimate for the energy contributions
$\Delta_{\text{excl}}(\kappa_{\min})$ of the excluded determinants $\ket{\Phi_{\nu}}$, i.e. those
determinants $\ket{\Phi_{\nu}}$ for which $|\kappa_{\nu}|<\kappa_{\min}$. In the limit of
a vanishing threshold, which means that no determinants are discarded, we obviously
have $\Delta_{\text{excl}}(0)=0$. We employ this property as a constraint in a simultaneous
extrapolation of the energies $E(\kappa_{\min})$ obtained from the truncated eigenvalue
problem and of the perturbatively corrected energies
$E(\kappa_{\min})+\Delta_{\text{excl}}(\kappa_{\min})$ using 5th to 7th order polynomials
for each of the quantities. The details of this procedure will be discussed elsewhere \cite{Roth08b}.
The uncertainties of the extrapolated energies reported in Sec. \ref{sec:benchmark} are always below 0.5 MeV. 

The process of constructing the importance-truncated model spaces, combined with
the extrapolation to the $\kappa_{\min}\to0$ limit, as described above,
is very efficient in reducing the dimensions of the corresponding CI or NCSM spaces
to the many-body basis states that are most relevant for the eigenvalue problem of interest.
Nevertheless, we still have to solve
relatively large eigenvalue problems. Since the resulting eigenvalue problems involve sparse
Hamiltonian matrices, we can use the Lanczos or Arnoldi technique to solve them. We use the
ARPACK library for that purpose. The largest dimensions of the importance-truncated model
spaces that appear in the calculations for \elem{O}{16} presented in Sec. \ref{sec:benchmark}
are a factor of a few times $10^7$. Since we solve the eigenvalue problem in a restricted space,
the importance-truncated CI and NCSM calculations fulfill the variational principle
and the Hylleraas-Undheim theorem \cite{HyUn30}. Furthermore, as in all CI and NCSM calculations,
the IT-CI and IT-NCSM approaches provide us with easy access to the
eigenstates of the Hamiltonian in a shell-model representation, along with the energies,
without any additional effort. The IT-CI and IT-NCSM eigenstates
can readily be used to compute expectation values of various observables,
transition matrix elements, or densities and form-factors. 

If we stop the iterative construction of the importance-truncated model
space before full self-consistency is reached, e.g. after just two iterations, as is done in this work,
then the resulting IT-CI(4p4h) approach is not strictly size extensive. Thus,
as in multi-reference CI calculations in quantum chemistry, where a similar problem occurs,
we employ the aforementioned multi-reference Davidson (MRD) correction in order to estimate the effect
of higher-order configurations beyond the 4p4h excitation level and to approximately
restore size extensivity. We use the multi-reference rather than single-reference
Davidson correction, since, in analogy to the multi-reference CI methods of quantum chemistry,
the IT-CI(4p4h) wave function expansions contain subsets of the 3p3h and 4p4h determinants
resulting from the up to 2p2h excitations from the multi-configurational reference state
$\ket{\Psi_\refe}$. There are several ways of calculating the MRD extensivity corrections
\cite{mrdci,sqdc1,sqdc2,sqdc3,sqdc4,DuDi94}. We use the Davidson-Silver form of
the MRD correction discussed in Ref. \cite{DuDi94}, which we add to the IT-CI(4p4h) energy to
obtain the final, approximately size extensive IT-CI(4p4h)+MRD result and which is calculated
using the energy of the multi-configurational reference state $\ket{\Psi_\refe}$
and the summed weight of the reference determinants in the IT-CI(4p4h) eigenstate.
By employing the {\it a posteriori} MRD corrections to IT-CI(4p4h) energies,
we can incorporate the effects of the most essential higher-than-4p4h configurations
in a computationally efficient manner without dealing with the higher-than-4p4h excitations explicitly.
In the coupled-cluster methods discussed in Sec. \ref{sec:cc} such contributions are represented
by the disconnected product terms involving one- and two-body clusters. As shown in this work,
the approximately size extensive IT-CI(4p4h)+MRD approach is competitive
with the rigorously size extensive and accurate
coupled-cluster schemes that are discussed in the following section.

\section{Coupled-Cluster Method}
\label{sec:cc}

The second class of methods that we consider are those based on the
single-reference coupled-cluster (CC) theory \cite{coester,coester2,cizek,cizek2,cizek3}
(see Refs. \cite{phys_rep_1978,ccphysrev2,ccreviews_bartlett,ccreviews_paldusli,ccreviews_schaefer,irpc:2002,PP:TCA}
for selected reviews), which utilizes the exponential ansatz for the $A$-particle ground state,
\beq
|\Psi_{0}\rangle={\exp}(T)|\Phi_{0}\rangle ,
\label{eq:cc}
\eeq
where $|\Phi_{0}\rangle$ is the reference determinant
and
\beq
T = \sum_{k=1}^{A} T_{k}
\label{eq:tdef}
\eeq
is the cluster operator. The cluster operator $T$ is a particle-hole
excitation operator, defined relative to the Fermi vacuum $|\Phi_{0}\rangle$, whose many-body
components
\beq
T_{k} = \sum_{i_{1} < \cdots < i_{k},a_{1} < \cdots < a_{k}}
t_{a_{1}\ldots a_{k}}^{i_{1} \ldots i_{k}} \,
a_{a_{1}}^{\dagger} \cdots a_{a_{k}}^{\dagger} a_{i_{k}} \cdots a_{i_{1}}
\label{eq:tk}
\eeq
generate the connected wave function diagrams of $|\Psi_{0}\rangle$ to infinite order.
The remaining linked, but disconnected contributions are produced through the exponential
ansatz for $|\Psi_{0}\rangle$. Here and elsewhere in this paper, we use the notation in which
$i_{1}, \: i_{2}, \ldots$ or $i, \: j, \ldots$ are the single-particle states
occupied in the Fermi vacuum state $|\Phi_{0}\rangle$ and
$a_{1}, \: a_{2}, \ldots$ or $a, \: b, \ldots$ are the single-particle states
unoccupied in $|\Phi_{0}\rangle$ (we will use labels $p,q,\ldots$ for the generic
single-particle basis states).

Typically, the explicit equations for the ground-state energy $E_{0}$ and the cluster amplitudes
$t_{a_{1}\ldots a_{k}}^{i_{1} \ldots i_{k}}$ defining the many-body components $T_{k}$ of $T$ according
to Eq. (\ref{eq:tk})
are obtained by inserting the wave function $|\Psi_{0}\rangle$, Eq. (\ref{eq:cc}),
into the Schr{\" o}dinger equation, $H |\Psi_{0}\rangle = E_{0} |\Psi_{0}\rangle$,
premultiplying both sides of the resulting equation on the left by $\exp(-T)$ to obtain
the connected cluster form of the Schr{\" o}dinger equation
\cite{cizek,cizek2},
\beq
\bar{H} |\Phi_{0} \rangle = E_{0} |\Phi_{0} \rangle ,
\label{rightcc}
\eeq
where
\beq
\bar{H}={\exp}(-T)H{\exp}(T) = [H \exp(T)]_{C}
\label{similaritycc}
\eeq
is the similarity-transformed Hamiltonian
or, equivalently, the connected product of the Hamiltonian and $\exp(T)$
(designated by subscript $C$),
and projecting both sides of Eq. (\ref{rightcc}) on the reference determinant $|\Phi_{0}\rangle$ and
excited determinants $|\Phi_{i_{1} \ldots i_{k}}^{a_{1} \ldots a_{k}} \rangle
= a_{a_{1}}^{\dagger} \cdots a_{a_{k}}^{\dagger} a_{i_{k}} \cdots a_{i_{1}} |\Phi_{0}\rangle$
that span the relevant $A$-particle Hilbert space. The latter projections result in a nonlinear
system of explicitly connected and energy-independent equations for the cluster amplitudes
$t_{a_{1}\ldots a_{k}}^{i_{1} \ldots i_{k}}$,
\beq
\langle \Phi_{i_{1} \ldots i_{k}}^{a_{1} \ldots a_{k}} |
\bar{H}|\Phi_{0}\rangle = 0 , \;\; i_{1}< \cdots < i_{k}, \;\;
a_{1} < \cdots < a_{k},
\label{cceqs}
\eeq
where $\bar{H}$ is defined by Eq. (\ref{similaritycc}) and $k=1,\ldots,A$, whereas the projection
on $|\Phi_{0}\rangle$ results in the CC energy formula,
\beq
E_{0} = \langle\Phi_{0}|\bar{H}|\Phi_{0}\rangle .
\label{ccenergy}
\eeq
The advantage of this formulation of CC theory, which is a standard
formulation adopted, for example, in quantum chemistry, where most of the
development and application work involving CC methods has occured, is
that, unlike the expectation value of the Hamiltonian with the CC wave function,
which would lead to a non-terminating power series in $T$ of the form
$E_{0} = \langle\Phi_{0}| [\exp(T^{\dagger}) H \exp(T)]_{C} |\Phi_{0}\rangle$ \cite{cizek2}, the CC equations
written above represent algebraic expressions that mathematically terminate at a finite power of $T$.
The similarity-transformed Hamiltonian $\bar{H}$, or the connected product of
the Hamiltonian and the ${\exp}(T)$ operator which is equivalent to $\bar{H}$ [cf. Eq. (\ref{similaritycc})],
is a finite polynomial expansion in $T$ whose
length depends only on the highest many-body rank of the interactions in the
Hamiltonian, not on the number of particles in a system,
so one does not need to make {\it ad hoc} truncations in powers of $T$,
which would have to be invoked if one attempted to
minimize the expectation value of the Hamiltonian,
in order to solve for the cluster amplitudes and energy.
For example,
for Hamiltonians that do not contain higher-than-pairwise interactions,
$\bar{H}$ and the resulting CC equations for cluster amplitudes, Eq. (\ref{cceqs}),
terminate at the $T^{4}$ terms, which is a purely mathematical truncation
resulting from the fact that one cannot connect more than four $T$ diagrams
to the diagrams representing the Hamiltonian in the definition of $\bar{H}$, Eq. (\ref{similaritycc}),
if $H$ is a two-body Hamiltonian. The energy formula, Eq. (\ref{ccenergy}),
simplifies even further in this case. If H does not contain higher-than-pairwise interactions,
we obtain
\beq
E_{0} = \langle \Phi_{0} | H | \Phi_{0} \rangle +
\langle \Phi_{0} | [ H_{N} (T_{1} + T_{2} + \half T_{1}^{2}) ]_{C} | \Phi_{0} \rangle ,
\label{energytwobody}
\eeq
where $H_{N} = H - \langle \Phi_{0} | H | \Phi_{0} \rangle$ is the Hamiltonian
in the normal-ordered form relative to $| \Phi_{0} \rangle$. In other words,
after determining the cluster operator $T$ by solving the system of equations
represented by Eq. (\ref{cceqs}),
we only need the 1p1h or singly excited and 2p2h or doubly excited
components of $T$, $T_{1}$ and $T_{2}$, respectively,
to determine the energy if $H$ is a two-body Hamiltonian.

The above is the exact CC theory, which is equivalent to the exact diagonalization
of the Hamiltonian with the full CI approach. Indeed, we could obtain
Eqs. (\ref{cceqs}) and (\ref{ccenergy}) by directly projecting the
Schr{\" o}dinger equation for the exact CC wave function $| \Psi_{0} \rangle$, Eq. (\ref{eq:cc}),
$H \exp(T) | \Phi_{0} \rangle  = E_{0} \exp(T) | \Phi_{0} \rangle$, with $T$ defined by
Eq. (\ref{eq:tdef}), on the
$|\Phi_{0}\rangle$ and $|\Phi_{i_{1} \ldots i_{k}}^{a_{1} \ldots a_{k}} \rangle$
determinants that span the $A$-particle Hilbert space. The energy-dependent terms on the right-hand side
of the resulting equations,
\beq
\langle \Phi_{i_{1} \ldots i_{k}}^{a_{1} \ldots a_{k}}| H \exp(T) | \Phi_{0} \rangle  = E_{0}
\langle \Phi_{i_{1} \ldots i_{k}}^{a_{1} \ldots a_{k}} | \exp(T) | \Phi_{0} \rangle ,
\label{cceqs-unlinked}
\eeq
cancel out the unlinked terms on the left-hand side of Eq. (\ref{cceqs-unlinked}) to produce
the system of the explicitly connected amplitude equations represented by Eq. (\ref{cceqs})
(see, e.g., Ref. \cite{leszcz} for a pedagogical derivation). Similarly, the disconnected
diagrams corresponding to the product of the Hamiltonian and $\exp(T)$ in the resulting energy formula,
\beq
E_{0} = \langle \Phi_{0} | H \exp(T) | \Phi_{0} \rangle ,
\label{ccenergy-dc}
\eeq
do not contribute to Eq. (\ref{ccenergy-dc}), resulting in Eq. (\ref{ccenergy}). Alternatively,
one can show that Eq. (\ref{ccenergy}) is equivalent to the expectation value of the Hamiltonian with
the CC wave function,
$\langle\Phi_{0}| \exp(T^{\dagger})H\exp(T) |\Phi_{0}\rangle/
\langle\Phi_{0}| \exp(T^{\dagger})\exp(T) |\Phi_{0}\rangle$,
as long as the cluster operator $T$ has the exact
form given by Eq. (\ref{eq:tdef}), in which all many-body components of $T$ including $T_{A}$ are included, and
as long as the corresponding cluster amplitudes $t_{a_{1}\ldots a_{k}}^{i_{1} \ldots i_{k}}$ with
$k=1,\ldots,A$ satisfy Eq. (\ref{cceqs}) (see, e.g., Ref. \cite{monk}).

The exact CC theory, as described above, being equivalent to the full CI diagonalization, is
limited to small
few-body problems. Thus, in all practical applications of CC theory, including those
reported in this work, one truncates the many-body expansion for $T$ at some,
preferably low, $m$p$m$h excitation level $T_{m}$. In this study, following the footsteps
of quantum chemistry, where the CC theory has become one of the most successful and
frequently used many-body methodologies, we focus on the most practical CC approximations
that can be applied to systems with dozens or even hundreds of correlated fermions. Thus,
we consider the basic CCSD (CC singles and doubles) approximation \cite{ccsd1,ccsd2,ccsdfritz,osaccsd},
which accounts for the effects of one- and two-body clusters, $T_{1}$ and $T_{2}$, respectively,
as well as the completely renormalized
CR-CC(2,3) approach \cite{crccl_jcp,crccl_cpl,crccl_open}, which accounts for
the effects of connected three-body $T_{3}$ clusters through a relatively inexpensive,
yet very effective non-iterative correction to the CCSD energy and which represents an
improved variant of the completely renormalized CCSD(T) [CR-CCSD(T)] method
\cite{irpc:2002,PP:TCA,leszcz,crcc_jcp} used in previous {\it ab initio} no core calculations
for the $^{4}$He and $^{16}$O nuclei \cite{kowalski04,o16prl,epja_2005,jpg_2005}.
In addition to many successful quantum chemistry applications (see Refs.
\cite{crccl_jcp,crccl_cpl,crccl_open,crcc23a,crcc23b,crcc23c,crcc23d,crcc23e,%
crcc23f,crcc23g,crcc23h,crcc23i,crcc23j} for representative examples), the CR-CC(2,3) approach
was recently used to study the ground and excited states of the $^{56}$Ni nucleus,
treated by the effective Hamiltonian in the $pf$-shell basis \cite{ni56_2007},
demonstrating its ability to provide a virtually exact description of
closed-shell nuclei at a tiny fraction of the costs of the shell-model calculations
that aim at similar accuracies, such as the CI method with up to 4p4h excitations,
and the ability to retain the accuracy of the CI approach with up to 4p4h excitations
even when the reference determinant contributes very little to the wave function
[the CPU timings of the CR-CC(2,3) calculations reported in Ref. \cite{ni56_2007}
were on the order of one minute on a single processor].
The present paper shows the high accuracy offered by the CR-CC(2,3) approach in the
context of the much larger scale no-core {\it ab initio} calculations for $^{16}$O
employing realistic nucleon-nucleon interactions and single-particle bases as large
as 8 major oscillator shells (480 uncoupled single-particle basis states).

In the CCSD calculations, which, in addition to producing reasonably accurate results
for closed-shell systems, provide the framework for the determination
of the CR-CC(2,3) corrections due to $T_{3}$ clusters, the
many-body expansion of the cluster operator $T$ defining
the CC wave function ansatz is truncated at the level of 2p2h (or double) 
excitations, i.e. $T=T_{1}+T_{2}$, where [cf. Eq. (\ref{eq:tk})]
\beq
  T_{1} 
  = \sum_{i,a}t_{a}^{i}a_{a}^{\dagger}a_{i}
\label{eq:t1}
\eeq
and 
\beq
  T_{2}
  = \sum_{i<j,a<b}t_{ab}^{ij}a_{a}^{\dagger}a_{b}^{\dagger}a_{j}a_{i}
\label{eq:t2}
\eeq
are the singly and doubly excited clusters.
In analogy to the exact CC theory defined by Eqs. (\ref{cceqs}), where $k=1,\ldots,A$, and (\ref{ccenergy}),
we determine the cluster amplitudes $t_{a}^{i}$ and $t_{ab}^{ij}$ defining the
CCSD wave function
\beq
|\Psi_{0}^{\rm (CCSD)}\rangle = \exp(T_{1}+T_{2}) |\Phi_{0} \rangle
\label{rightccsd}
\eeq
by solving the
nonlinear system of energy-independent and explicitly connected algebraic equations
similar to Eq. (\ref{cceqs}). Specifically, since we only need the $T_{1}$ and $T_{2}$
cluster components, we consider the subset of equations corresponding to the
projections on the singly and doubly excited determinants [Eq. (\ref{cceqs}) with $k=1$ and 2],
so that the number of equations matches the number of unknown amplitudes $t_{a}^{i}$ and $t_{ab}^{ij}$.
We obtain
\beq
  \langle\Phi_{i}^{a}|\bar{H}^{\rm (CCSD)}|\Phi_{0}\rangle=0 ,
\label{ccsd:eqt1}
\eeq
\beq
  \langle\Phi_{ij}^{ab}|\bar{H}^{\rm (CCSD)}|\Phi_{0}\rangle=0 ,
\label{ccsd:eqt2}
\eeq
where 
\begin{eqnarray}
  \bar{H}^{\rm (CCSD)} & = & {\exp}(-T_{1}-T_{2})H{\exp}(T_{1}+T_{2})
  \nonumber \\
  & = & [H \exp(T_{1}+T_{2})]_{C}
\label{hbarccsd}
\end{eqnarray}
is the similarity-transformed Hamiltonian of CCSD [the similarity-transformed Hamiltonian $\bar{H}$,
Eq. (\ref{similaritycc}), written for $T=T_{1}+T_{2}$], and $|\Phi_{i}^{a}\rangle$ and
$|\Phi_{ij}^{ab}\rangle$ are the singly and doubly excited determinants,
respectively, relative to $|\Phi_{0}\rangle$. Once the cluster amplitudes
$t_{a}^{i}$ and $t_{ab}^{ij}$ defining $T_{1}$ and $T_{2}$ are
known, the CCSD energy $E_{0}^{\rm (CCSD)}$ is computed using Eq. (\ref{ccenergy}) in which
$\bar{H}$ is replaced by $ \bar{H}^{\rm (CCSD)}$, Eq. (\ref{hbarccsd}).

As pointed out above, the CC equations obtained through projections of the
Schr{\" o}dinger equation have finite polynomial nature, i.e. they terminate at a finite
power of $T$ which depends on the highest many-body rank of the interactions in the Hamiltonian.
When the Hamiltonian is a two-body operator, the CC equations for cluster amplitudes
terminate at the $T^{4}$ terms and the ground-state energy is given by Eq. (\ref{energytwobody}).
In particular, when $H$ does not contain higher-than-two-body interactions,
the generic CCSD amplitude equations, Eqs. (\ref{ccsd:eqt1}) and (\ref{ccsd:eqt2}),
that apply to all Hamiltonians simplify to
\beq
\langle \Phi_{i}^{a} |
[H_{N}(1 + T_{1} + T_{2} + \half T_{1}^{2} + T_{1} T_{2}
+ \six T_{1}^{3} )]_{C}
|\Phi_{0} \rangle = 0
\label{momccsd1}
\eeq
and
\begin{eqnarray}
\langle \Phi_{ij}^{ab} |
[H_{N}(1 \!\!\!\!\! && + T_{1} + T_{2} + \half T_{1}^{2} + T_{1} T_{2}
+ \six T_{1}^{3} + \half T_{2}^{2}
\nonumber \\
&& + \half T_{1}^{2} T_{2}
+ \tfour T_{1}^{4} )]_{C}
|\Phi_{0} \rangle = 0 ,
\label{momccsd2}
\end{eqnarray}
respectively. If the cluster operator $T$ was not truncated, as in the exact CC theory defined by
Eqs. (\ref{eq:tdef}), (\ref{cceqs}), and (\ref{ccenergy}), the results would be equivalent to the variational,
full CI calculation corresponding to the minimization of the expectation value of the Hamiltonian with
the CC wave function $|\Psi_{0}\rangle$, Eq. (\ref{eq:cc}).
Unfortunately, in the CCSD case, $T$ is truncated at $T_{2}$, so that the
resulting ground-state energy $E_{0}^{\rm (CCSD)}$, determined by solving the CCSD amplitude equations,
Eqs. (\ref{ccsd:eqt1}) and (\ref{ccsd:eqt2}) [or,
when the Hamiltonian is two-body, (\ref{momccsd1}) and (\ref{momccsd2})], and by using
the resulting $T_{1}$ and $T_{2}$ clusters in Eq. (\ref{ccenergy}) [or, in the two-body Hamiltonian case,
(\ref{energytwobody})] is not equivalent to the expectation value of the Hamiltonian with the
CCSD wave function $|\Psi_{0}^{\rm (CCSD)}\rangle$, Eq. (\ref{rightccsd}), and, as such,
does not have the bound if $A > 2$. One could, at least in principle, contemplate variational CCSD
calculations based on minimizing the expectation value of the Hamiltonian with the CCSD wave function,
as in Refs. \cite{vcc1,vcc2}, but, as already explained, the expectation value of the Hamiltonian with the
CC (e.g. CCSD) wave function is a non-terminating series in cluster amplitudes that does not
lead to practical computational schemes. On the other hand,
due to the exponential nature of the CC wave function,
the approximate CC methods using
truncated cluster operators $T = \sum_{k=1}^{m} T_{k}$ with $m < A$ (CCSD corresponds to
the $m=2$ case)
converge more rapidly to the full CI limit as $m \rightarrow A$
than the equivalent truncated shell-model expansions that rely on the same manifold of
excited determinants, defined by Eq. (\ref{truncated-ci}), while providing the rigorously size extensive
results which truncated CI expansions cannot provide
due to the presence of unlinked contributions in the wave function and disconnected
contributions in the energy that do not cancel out.
For example, as already established in the 1970s and early 1980s,
the CCSD method, which relies on up to 2p2h cluster components,
is more accurate than the CI approach
truncated at 2p2h determinants. As mentioned earlier, and as demonstrated in this article,
the CCSD approach corrected for the connected 3p3h clusters $T_{3}$ via the CR-CC(2,3) method
is as accurate as the CI approach with up to 4p4h excitations, while providing a size extensive
description of many-particle systems. The issue of the lack of extensivity in CI calculations
is addressed in this paper through the suitably defined, quantum-chemistry inspired
corrections to the CI energies, as described in the previous section. The elementary analysis
that enables one to understand why the CR-CC(2,3) method
is as accurate as the CI approach with up to 4p4h excitations is provided in Section \ref{section4c}.

The CCSD approximation provides information about the bulk of the correlation effects,
which are described by $T_{1}$, $T_{2}$, and their products, but
one must account for the higher-than-doubly excited connected clusters, particularly
the $T_3$ clusters, in order to obtain a more quantitative description.
This can be done through the
CCSDT (CC singles, doubles, and triples) method 
\cite{ccsdt1,ccsdt2}, in which the
cluster operator is truncated at the 3p3h level ($T = T_{1}+T_{2} + T_{3}$) and, in addition
to the equations corresponding to the projections on singly and doubly excited determinants,
one considers the projections on triply excited determinants [$k=3$ in Eq. (\ref{cceqs})].
Unfortunately, as explained below,
the CCSDT scheme has very large computational costs, and so is generally not 
practical and limited to small few-body problems and relatively small single-particle basis sets. As a result,
inspired by the similar difficulties encountered in quantum chemistry, where CCSDT is limited to
small few-electron systems, we instead use the considerably less expensive CR-CC(2,3) approach, in which
we incorporate the effects of $T_3$ clusters by adding a non-iterative {\it a posteriori} correction 
\beq
  \delta_{0}(2,3)=\sum_{i<j<k,a<b<c}{\ell}_{ijk}^{abc} \, {\mathfrak M}_{abc}^{ijk}
\label{delta23}
\eeq
to the CCSD energy $E_{0}^{\rm (CCSD)}$, so that the final CR-CC(2,3) energy is
\beq
E_{0}^{\rm (CR\mbox{-}CC(2,3))} = E_{0}^{\rm (CCSD)} + \delta_{0}(2,3) ,
\label{crcc23}
\eeq
where $\delta_{0}(2,3)$ is given by Eq. (\ref{delta23}). The quantities
${\mathfrak M}_{abc}^{ijk}$ entering Eq. (\ref{delta23}) are the
projections of the connected cluster form of the Schr{\" o}dinger equation written
for the CCSD wave function on the triply excited determinants $|\Phi_{ijk}^{abc}\rangle$,
which define the
triply excited moments of the CCSD equations
\cite{irpc:2002,PP:TCA,leszcz,crccl_jcp,crccl_cpl,crccl_open,crcc_jcp}, i.e.
\beq
{\mathfrak M}_{abc}^{ijk}=\langle\Phi_{ijk}^{abc}|\bar{H}^{\rm (CCSD)}|\Phi_{0}\rangle ,
\label{mom3}
\eeq
where $\bar{H}^{\rm (CCSD)}$ is the similarity-transformed Hamiltonian of the CCSD approach,
Eq. (\ref{hbarccsd}). As in the case of the CCSD equations, the explicit 
form of Eq. (\ref{mom3}) depends on the nature of the interactions in the Hamiltonian. When
the Hamiltonian does not contain higher--than--two-body interactions (the case examined in this
paper), one can write
\begin{eqnarray}
{\mathfrak M}_{abc}^{ijk} & = & \langle \Phi_{ijk}^{abc} |
[H_{N}(T_{2} + T_{1} T_{2} + \half T_{2}^{2} + \half T_{1}^{2} T_{2}
\nonumber \\
&&
+ \half T_{1} T_{2}^{2} + \six T_{1}^{3} T_{2})]_{C}
|\Phi_{0} \rangle .
\label{momccsd3}
\end{eqnarray}
The ${\ell}_{ijk}^{abc}$ coefficients entering Eq. (\ref{delta23}) are defined as
\beq
{\ell}_{ijk}^{abc}=\langle\Phi_{0}|\Lambda\bar{H}|\Phi_{ijk}^{abc}\rangle/D_{abc}^{ijk} ,
\label{ellijkabc}
\eeq
where the denominator
\beq
D_{abc}^{ijk} = E_{0}^{\rm (CCSD)} - \langle\Phi_{ijk}^{abc}|\bar{H}^{\rm (CCSD)}|\Phi_{ijk}^{abc}\rangle
\eeq
is obtained by approximating the triples-triples block of the matrix representing the
$\bar{H}^{\rm (CCSD)}$ operator, Eq. (\ref{hbarccsd}),
by its diagonal, as in the Epstein-Nesbet perturbation theory,
while $\Lambda=\Lambda_{1}+\Lambda_{2}$ is the hole-particle deexcitation operator
defining the CCSD `bra' or dual ground state \cite{stanton1993,pprjb1999,aip2005}
\beq
\langle\tilde{\Psi}_{0}^{\rm (CCSD)}|=\langle\Phi_{0}|(1+\Lambda_{1}+\Lambda_{2}){\exp}(-T_{1}-T_{2}),
\label{leftccsd}
\eeq
which satisfies the biorthonormality condition
$\langle\tilde{\Psi}_{0}^{\rm (CCSD)}| \Psi_{0}^{\rm (CCSD)}\rangle = 1$. The one- and two-body
components of the $\Lambda$ operator of CCSD,
\beq
\Lambda_{1} = \sum_{i,a} \lambda_{i}^{a} a_{i}^{\dagger} a_{a}
\label{eq:lambda1}
\eeq
and
\beq
\Lambda_{2} = \sum_{i<j,a<b} \lambda_{ij}^{ab} a_{i}^{\dagger} a_{j}^{\dagger} a_{b} a_{a},
\label{eq:lambda2}
\eeq
respectively, are obtained by
solving the linear system of equations \cite{stanton1993,pprjb1999,aip2005}
\beq
\langle \Phi_{0}| (1+\Lambda_{1}+\Lambda_{2})\bar{H}^{\rm (CCSD)}|\Phi_{i}^{a}\rangle = E_{0}^{\rm (CCSD)} \lambda_{i}^{a} ,
\label{ccsd:eqlambda1}
\eeq
\beq
\langle \Phi_{0}| (1+\Lambda_{1}+\Lambda_{2})\bar{H}^{\rm (CCSD)}|\Phi_{ij}^{ab}\rangle = E_{0}^{\rm (CCSD)} \lambda_{ij}^{ab} .
\label{ccsd:eqlambda2}
\eeq
The details of the derivation of the non-iterative correction $\delta_{0}(2,3)$
defining the CR-CC(2,3) calculations,
which originates from the mathematical formalism referred to as the
biorthogonal method of moments of coupled-cluster equations
\cite{crccl_jcp,crccl_cpl} and which describes the leading terms toward the
full CI energy due to $T_3$ clusters, mimicking the performance of
the much more expensive CCSDT method at the small fraction of the computer cost,
can be found in Refs. \cite{crccl_jcp,crccl_cpl}.

The fact that the construction of the CR-CC(2,3) correction $\delta_{0}(2,3)$
requires the determination of the $\Lambda$ operator of the CCSD theory, which defines the
CCSD bra state $\langle\tilde{\Psi}_{0}^{\rm (CCSD)}|$, as described above,
has an additional advantage that we can use the same operator $\Lambda$ to determine
properties other than energy. We can, for example, use it to determine the CCSD one-body
reduced density matrices 
\begin{eqnarray}
  \gamma_{q}^{p} 
  & \equiv & \langle\tilde{\Psi}_{0}^{\rm (CCSD)}| (a_{p}^{\dagger} a_{q}) | \Psi_{0}^{\rm (CCSD)} \rangle 
\nonumber \\
  & = & \langle \Phi_{0} | (1+\Lambda_{1}+\Lambda_{2}) \overline{(a_{p}^{\dagger} a_{q})} | \Phi_{0} \rangle
\label{ccsd-one-rdm}
\end{eqnarray}
and properties
\beq
  \langle\tilde{\Psi}_{0}^{\rm (CCSD)}| \Theta | \Psi_{0}^{\rm (CCSD)} \rangle 
  = \sum_{p,q} \theta_{p}^{q} \, \gamma_{q}^{p} ,
\label{ccsd-property}
\eeq
where $\Theta$ is a one-body property operator defined through matrix elements
$\theta_{p}^{q} \equiv \langle p | \theta | q \rangle$,
\begin{eqnarray}
  \overline{(a_{p}^{\dagger} a_{q})} & = & {\exp}(-T_{1}-T_{2}) (a_{p}^{\dagger} a_{q}) {\exp}(T_{1}+T_{2})
\nonumber \\
  & = & [(a_{p}^{\dagger} a_{q}) \exp(T_{1}+T_{2})]_{C}
\label{ccsd-adagger-a}
\end{eqnarray}
is the similarity-transformed connected form of the operator
string $(a_{p}^{\dagger} a_{q})$, analogous to the similarity-transformed Hamiltonian $\bar{H}^{\rm (CCSD)}$,
$T_{1}$ and $T_{2}$ are the singly and doubly excited clusters obtained by solving the CCSD equations,
Eqs. (\ref{ccsd:eqt1}) and (\ref{ccsd:eqt2}), and $\Lambda_{1}$ and $\Lambda_{2}$ are obtained by solving the
linear system of equations given by Eqs. (\ref{ccsd:eqlambda1}) and (\ref{ccsd:eqlambda2})
\cite{stanton1993,pprjb1999,aip2005}. In general, if $\Xi$ is an operator representing the quantum-mechanical
quantity of interest, the CC analog of the expectation value of $\Xi$ can be determined using the equation
\cite{stanton1993,pprjb1999,aip2005}
\beq
\langle \Xi \rangle \equiv  \langle\tilde{\Psi}_{0}| \Xi | \Psi_{0} \rangle = \langle \Phi_{0} | (1+\Lambda) \,
\bar{\Xi} | \Phi_{0} \rangle,
\label{cc-property}
\eeq
where
\beq
\bar{\Xi} = {\exp}(-T) \, \Xi \, {\exp}(T) = [\Xi \exp(T)]_{C}
\label{similarity-property}
\eeq
is the similarity-transformed form of $\Xi$ and $\Lambda$ is the hole-particle deexcitation operator defining
the left ground eigenstate of $\bar{H}$, Eq. (\ref{similaritycc}), which has the form
$\langle \Phi_{0} | (1+\Lambda)$, and the bra or dual CC ground state
\beq
\langle\tilde{\Psi}_{0}|=\langle\Phi_{0}|(1+\Lambda){\exp}(-T) .
\label{leftcc}
\eeq
In analogy to the CCSD case, $\Lambda$
can be obtained by solving the linear system of equations,
\beq
\langle \Phi_{0} | (1+\Lambda) \bar{H} |
\Phi_{i_{1} \ldots i_{k}}^{a_{1} \ldots a_{k}} \rangle = E_{0}
\lambda^{a_{1}\ldots a_{k}}_{i_{1} \ldots i_{k}}  ,
\label{cceqs-lambda}
\eeq
where $E_{0}$ is the CC ground-state energy and
$\lambda^{a_{1}\ldots a_{k}}_{i_{1} \ldots i_{k}}$ are the amplitudes that define the many-body components of $\Lambda$,
\beq
\Lambda_{k} =
\sum_{
i_{1} < \cdots < i_{k},
a_{1} < \cdots < a_{k}
}
\lambda^{a_{1}\ldots a_{k}}_{i_{1} \ldots i_{k}} \:
a_{i_{1}}^{\dagger} \cdots a_{i_{k}}^{\dagger} a_{a_{k}} \cdots a_{a_{1}}
\label{eq:lambdak}
\eeq
[clearly, Eq. (\ref{cceqs-lambda}) is the generalization of Eqs. (\ref{ccsd:eqlambda1})
and (\ref{ccsd:eqlambda2}), corresponding to the CCSD case, to any
level of CC theory].
In particular, when $\Xi$ is the Hamiltonian $H$, we obtain from Eq. (\ref{cc-property})
\beq
\langle H \rangle = \langle \Phi_{0} | (1+\Lambda)
\bar{H} | \Phi_{0} \rangle,
\label{hamiltonian-expectation}
\eeq
which is equivalent to the CC energy formula, Eq. (\ref{ccenergy}), since the
$\langle \Phi_{0} | \Lambda_{k} \bar{H} | \Phi_{0} \rangle$ contributions to Eq. (\ref{hamiltonian-expectation})
vanish when the cluster operator $T$ entering the definition of $\bar{H}$, as in Eq.
(\ref{similaritycc}), satisfies the system of CC equations, Eq. (\ref{cceqs}).
It should be noted that the above way of calculating the $\langle \Xi \rangle$ values, which
reflects on the biorthogonal character of CC theory, becomes fully equivalent to the determination
of $\langle \Xi \rangle$ as the conventional expectation value
$\langle \Psi_{0} | \Xi | \Psi_{0} \rangle/\langle \Psi_{0} | \Psi_{0} \rangle$ when $T$ is
a non-truncated cluster operator given by Eq. (\ref{eq:tdef}) obtained in the exact CC calculations
defined by Eq. (\ref{cceqs}). When $T$ is truncated, as in the CCSD case, the value of
$\langle \Xi \rangle$ determined from Eq. (\ref{cc-property}), although no longer identical to the
quantum-mechanical expectation value
$\langle \Psi_{0} | \Xi | \Psi_{0} \rangle/\langle \Psi_{0} | \Psi_{0} \rangle$, where $| \Psi_{0} \rangle$
is the corresponding CC wave function, is equivalent to the
alternative way of determining $\langle \Xi \rangle$ as $(\partial E_{0}(\lambda)/\partial \lambda)_{\lambda=0}$,
where $E_{0}(\lambda)$ is the CC energy, Eq. (\ref{ccenergy}), calculated after solving the relevant CC
equations for the auxiliary Hamiltonian $H_{\lambda} = H + \lambda \Xi$, as in the response CC theory
\cite{monk,kondo-piecuch-paldus-lrcc}, as long as the reference determinant $|\Phi_{0}\rangle$ does not
vary with $\lambda$. We use this fact in Section \ref{section4a} to determine the expectation values of the
center-of-mass Hamiltonian corresponding to the CCSD and CR-CC(2,3) calculations as the appropriate CC
energy derivatives. The vast experience with performing CC calculations in quantum chemistry is that the
difference between the values of $\langle \Xi \rangle$ calculated as the conventional quantum-mechanical
expectation values and the $\langle \Xi \rangle$ values obtained as the corresponding energy derivatives,
as described above, are very small, since the approximate CC methods, such as those used in this work,
provide results close to full CI. Clearly, there are no differences between the $\langle \Xi \rangle$
values obtained as the traditional expectation values and energy derivatives in truncated CI calculations, since
all CI methods are variational and, as such, satisfy the Hellmann-Feynman theorem.
In approximate CC methods, we have to rely on a response formulation and the equations such as Eq. (\ref{cc-property}),
or the equivalent energy derivatives, as described above, since, in analogy to the Hamiltonian, the
traditional expectation value expression with the CC wave function,
$\langle \Psi_{0} | \Xi | \Psi_{0} \rangle/\langle \Psi_{0} | \Psi_{0} \rangle$,
would lead to a non-terminating power series in cluster amplitudes of the form
$\langle\Phi_{0}| [\exp(T^{\dagger}) \, \Xi \, \exp(T)]_{C} |\Phi_{0}\rangle$ that does not
lead to practical computational schemes.

By using the CR-CC(2,3) approach, we are able to significantly reduce the computational costs
associated with the inclusion of the connected triply excited clusters, enabling calculations
with relatively large model spaces, while producing results that
should be as accurate as those obtained with the full CCSDT scheme \cite{crccl_jcp}.
Indeed, the most expensive computational steps of CR-CC(2,3)
scale as $n_{o}^{3}n_{u}^{4}$ in the determination of the non-iterative correction due to $T_3$ and
$n_{o}^{2}n_{u}^{4}$ in the underlying CCSD calculation, where, as mentioned earlier,
$n_{o}$ and $n_{u}$ represent the numbers
of occupied and unoccupied single-particle states, respectively. For realistic values of $n_{o}$ and $n_{u}$,
including the larger single-particle basis sets used in this work, this is
less expensive than the $n_{o}^{3}n_{u}^{5}$ iterative steps defining CCSDT by orders of magnitude.
In addition, unlike in CCSDT, in the CR-CC(2,3) calculations
one does not have to store the large number of $\sim n_{o}^{3} n_{u}^{3}$ amplitudes $t_{abc}^{ijk}$ defining
the $T_{3}$ cluster operator, since the non-iterative correction $\delta_{0}(2,3)$
defining the $T_{3}$ correction to the CCSD energy, Eq. (\ref{delta23}), is calculated using the
one- and two-body clusters, $T_{1}$ and $T_{2}$, and their $\Lambda$ deexcitation analogs,
$\Lambda_{1}$ and $\Lambda_{2}$, respectively, as described above, which need a storage of
the $n_{o} n_{u}$ amplitudes $t_{a}^{i}$ and $\lambda_{i}^{a}$ and $n_{o}^{2} n_{u}^{2}$
amplitudes $t_{ab}^{ij}$ and $\lambda_{ij}^{ab}$. Thus, the total storage
requirements of the CR-CC(2,3)
calculations are primarily defined by the number of two-body matrix elements that define the Hamiltonian
or the two-body matrix elements of the similarity-transformed Hamiltonian
$\bar{H}^{\rm (CCSD)}$. The memory requirements
scale as $n_{o} n_{u}^{3}$. A small subset of the three-body matrix
elements of $\bar{H}^{\rm (CCSD)}$ that enter the calculation of
the CR-CC(2,3) triples correction $\delta_{0}(2,3)$ via the
aforementioned equations
for the $\Lambda$ operator of CCSD and the triply excited moments ${\mathfrak M}_{abc}^{ijk}$
do not have to be precomputed and stored, since one can rigorously
factorize the diagrams that represent them and express all quantities that enter
the calculation of $\delta_{0}(2,3)$ in terms of
the one and two-body matrix elements of $\bar{H}^{\rm (CCSD)}$ (see, e.g., Ref. \cite{wloch2005}, and references therein).
Thanks to the consistent use of the recursively generated intermediates,
including one- and two-body matrix elements
of $\bar{H}^{\rm (CCSD)}$, and fast matrix multiplication routines \cite{sak1}, as is always done in the most
efficient, modern implementations
of CC methods in quantum chemistry,
our CCSD and CR-CC(2,3) computer codes are fully vectorized. Thanks to the DIIS algorithm
\cite{pulay1}, which we use to solve the CCSD equations and their $\Lambda$
counterparts (for the first application
of the DIIS procedure to solving the CC equations, see Ref. \cite{ccdiis}),
our CCSD calculations typically converge
in about 20 iterations to obtain an energy accurate to within $10^{-5}$ MeV.
The triples correction $\delta_{0}(2,3)$
requires no iterations, which is one of the biggest advantages of CR-CC(2,3)
as opposed to the iterative CCSDT approach.
We refer the reader to Refs.
\cite{crccl_jcp,crccl_cpl,crccl_open} for the details of the CR-CC(2,3) theory, Ref. \cite{wloch2005} for the
explicit, computationally efficient
expressions for the one- and two-body matrix elements of $\bar{H}^{\rm (CCSD)}$ and moments ${\mathfrak M}_{abc}^{ijk}$ that
enter the CR-CC(2,3) calculations, and Ref. \cite{prc2006} for the computationally efficient,
fully factorized form of the CCSD equations that precede the calculation of
the triples correction $\delta_{0}(2,3)$.
We also note that while 
the earlier form of the CR-CC(2,3) approach, designated as CR-CCSD(T) \cite{irpc:2002,PP:TCA,leszcz,crcc_jcp},
which was used in the studies of
the $^{4}$He and $^{16}$O nuclei \cite{kowalski04,o16prl}, was only approximately size extensive
(to within 0.5--1 {\%} of the correlation energy \cite{irpc:2002}), the CR-CC(2,3) method used in this work is
fully size extensive, so that no loss of accuracy occurs when going from smaller to larger many-fermion systems.

One of the most important findings of this work is that the CR-CC(2,3) method, with its relatively
inexpensive computational steps that scale as $n_{o}^{2} n_{u}^{4}$ in the iterative part
and $n_{o}^{3} n_{u}^{4}$ in the non-iterative part,
is capable of providing the results that are virtually identical to those that effectively
correspond to the diagonalization of the Hamiltonian using the CI
approach with up to 4p4h excitations from the reference determinant $|\Phi_{0}\rangle$, corrected
for the effects of higher-than-4p4h excitations and
extensivity through the use of the Davidson corrections.
The IT-CI(4p4h) approach reduces the most expensive $n_{o}^{4} n_{u}^{6}$ steps of the full CI(4p4h)
calculation and the need to deal with the large numbers of the 3p3h and 4p4h determinants
in the CI(4p4h) diagonalization by orders of magnitude through the use of importance truncation
and extrapolation to the $\kappa_{\min}\to0$ limit, as discussed in the previous section.
The CR-CC(2,3) method does effectively the same work,
eliminating, in particular, the need to deal with the 3p3h and 4p4h excitations of CI in an explicit manner,
by representing the dominant higher-order correlations through the
disconnected product terms involving low-order $T_{1}$ and $T_{2}$ clusters,
such as, for example, $T_{1} T_{2}$ for 3p3h excitations, $(1/2)T_{2}^{2}$ for 4p4h excitations,
$(1/2) T_{1} T_{2}^{2}$ for 5p5h excitations, $(1/6) T_{2}^{3}$ for 6p6h excitations, etc.,
while making sure that the connected 3p3h excitations that are usually more important than
the disconnected $T_{1} T_{2}$ terms are accounted for through the computationally affordable
non-iterative $\delta_{0}(2,3)$ correction to the CCSD energy.
The detailed analysis of the CI and CC wave functions that explains this is provided in Sec. \ref{section4c}.
The numerical similarity of the CR-CC(2,3) and Davidson-corrected IT-CI(4p4h) [i.e.
IT-CI(4p4h)+MRD)] results does not address the issue of the center-of-mass contaminations
which are present in all truncated CC and all truncated CI calculations other than
NCSM. The impact of the center-of-mass contaminations on the truncated CC [CCSD and CR-CC(2,3)]
and CI [IT-CI(4p4h)] results for \elem{O}{16} is examined in Section \ref{section4a}.

\section{Benchmark for ${}^{16}$O}
\label{sec:benchmark}

We aim at a quantitative comparison of the different approaches using the ground state of \elem{O}{16} as an example. For the interaction we use the realistic $\VO_{\UCOM}$ potential derived in the framework of the Unitary Correlation Operator Method (UCOM) discussed in Refs. \cite{FeNe98,NeFe03,RoNe04,RoHe05}. Using a unitary transformation tailored for the description of short-range central and tensor correlations, a phase-shift equivalent soft interaction is derived from the Argonne V18 potential \cite{WiSt95}. Applications in various many-body approaches, from the no-core shell model for light nuclei to Hartree-Fock and many-body perturbation theory for heavy nuclei \cite{RoPa06}, show that this two-body potential allows for a realistic description of binding energies throughout the nuclear chart without the explicit inclusion of additional three-body (or other higher-than-two-body)
interactions. All calculations presented here are based on the same $\VO_{\UCOM}(I_{\vartheta}=0.09\fm^3)$ potential
as that used in Refs. \cite{RoHe05, RoPa06}.

As implied by the discussion in the previous section, we compare the following three groups of {\it ab initio} methods: (i) the coupled-cluster approach including singly and doubly excited clusters, CCSD, as well as the non-iterative completely renormalized CR-CC(2,3) scheme that corrects the CCSD results for the effects of the connected triply excited clusters, (ii) the importance-truncated configuration interaction method including up to 4p4h excitations, IT-CI(4p4h), without and with the MRD size extensivity corrections, and (iii) the importance-truncated no-core shell model approach including up to 4p4h excitations, IT-NCSM(4p4h), with and without the MRD corrections. In addition to comparing the results of CC, CI, and NCSM calculations, particularly the CR-CC(2,3) and IT-CI(4p4h)(+MRD) levels, we examine several issues relevant for this comparison, including the role of center-of-mass contaminations and the sensitivity of the results to the choice of the single-particle basis. By comparing the size extensive, but not variationally bound CR-CC(2,3) results with the variational, but not rigorously size extensive IT-CI(4p4h)(+MRD) results for the binding energies of ${}^{16}$O side by side, we have an opportunity to comment on the significance or insignificance of such issues as the violation of size extensivity by the truncated CI calculations and the lack of the variational bound in the CC calculations.

We begin with two important issues relevant for the comparison of the CCSD, CR-CC(2,3), IT-CI(4p4h)(+MRD), and IT-NCSM(4p4h) results, namely, the role of center of mass contaminations and the sensitivity of the results to the choice of the single-particle basis. 

\subsection{Center-of-mass problem}
\label{section4a}

All calculations are performed with a translationally invariant intrinsic Hamiltonian $\HO_{\intr} =  \TO-\TO_{\cm}+\VO_{\UCOM}$. However, this does not imply that the resulting many-body states and intrinsic energies are free of spurious contributions induced by a coupling of intrinsic and center-of-mass (CM) degrees of freedom. For the nucleus as a self-bound and translationally invariant system the intrinsic properties should not depend on the CM motion, i.e. intrinsic and CM components of the many-body state have to decouple. 

For a Slater determinant basis, an exact separation of intrinsic and CM motions is possible only in a complete $N_{\max}\hbar\Omega$ model space based on harmonic oscillator single-particle states as employed in the full NCSM approach. The use of a different model-space truncation or a different single-particle basis destroys this decoupling property and induces CM contaminations. This problem is well known in the context of the nuclear shell model (see Ref. \cite{RaFa90} and references therein) and it was also addressed in the context of CC calculations for nuclei, both intrinsically, through a translationally invariant formulation \cite{MoWa02,BiFl90}, and numerically, through heuristic CM corrections added to the Hamiltonian \cite{bogdan1,bogdan2,bogdan4,kowalski04,o16prl,epja_2005,jpg_2005,prc2006,HaDe07}.

In order to probe to what extent intrinsic and CM motions are coupled in the different many-body
approaches examined in this work, we study the impact of an artificial shift of the CM spectrum using the Lawson prescription \cite{Palu67,GlLa74,RaFa90}. We introduce the modified Hamiltonian 
\eq{
  \HO_{\beta} = \HO_{\intr} + \beta \HO_{\cm} \;,
}
where 
\eq{
  \HO_{\cm} = \frac{1}{2Am}\POV^2_{\cm} + \frac{Am\Omega^2}{2} \XOV^2_{\cm} - \frac{3}{2}\hbar\Omega ,
}
with CM momentum $\POV_{\cm}$ and CM coordinate $\XOV_{\cm}$. If the CM motion is completely decoupled, then the expectation value of the intrinsic Hamiltonian $\HO_{\intr}$, $\expect{\HO_{\intr}}$, computed with the ground state resulting from the modified Hamiltonian $\HO_{\beta}$ is independent of the value of $\beta$. Any dependence of $\expect{\HO_{\intr}}$ on $\beta$ indicates an unphysical coupling of intrinsic and CM motions and a violation of translational invariance. 

\begin{table}
\caption{The CI and NCSM expectation values of $\HO_{\intr}$ and $\HO_{\cm}$, and their CC analogs defined in the text
(in units of MeV), obtained from the many-body solutions using  $\HO_{\beta}$ for $\beta=0$ and $\beta=10$. The CC and IT-CI calculations use a harmonic oscillator basis with $e_{\max}=5$ and $\hbar\Omega = 30\MeV$, whereas the IT-NCSM and NCSM calculations use a model space with $N_{\max}=8$ and $\hbar\Omega=22\MeV$. The oscillator frequencies correspond to the respective energy minima of the CC and NCSM calculations.}
\label{tab:cm_comparison}
\begin{ruledtabular}
\begin{tabular}{l c c c c}
 & \multicolumn{2}{c}{$\beta=0$} &   \multicolumn{2}{c}{$\beta=10$} \\
 & $\expect{\HO_{\intr}}$ &  $\expect{\HO_{\cm}}$  & $\expect{\HO_{\intr}}$ &  $\expect{\HO_{\cm}}$ \\
\hline
CCSD          & -107.32  & 5.88 & -104.84 & 0.24 \\
CR-CC(2,3)    & -113.14  & 5.38 & -111.23 & 0.20 \\
IT-CI(4p4h)   & -98.67   & 1.37 & -97.32  & 0.19 \\
\hline
IT-NCSM(4p4h) & -104.10  & 0.08  & -104.01 & 0.02 \\
NCSM          & -104.75  & 0.00  & -104.75 & 0.00  \\
\end{tabular}
\end{ruledtabular}
\end{table}

In Table \ref{tab:cm_comparison} we compare the results for the binding energy of \elem{O}{16}
obtained with the different many-body methods used in this work for two $\beta$ values,
namely, $\beta=0$, at which the initial intrinsic Hamiltonian is recovered, and $\beta=10$,
which is a typical empirical value used in shell model applications \cite{RaFa90}.
We use the representative $\hbar\Omega$ values which approximately match the
minima on the curves that describe the dependence of the relevant energies on $\hbar\Omega$.
Since there is no rigorous criterion for choosing $\beta$,
we will come back to the impact of variations of this parameter later on.
The IT-CI(4p4h), IT-NCSM(4p4h), and NCSM values of $\expect{\HO_{\intr}}$ and $\expect{\HO_{\cm}}$
are calculated directly as the standard quantum-mechanical expectation values of the relevant Hamiltonians using the
eigenvectors obtained from the solution of the eigenvalue problem of $\HO_{\beta}$.
The CCSD and CR-CC(2,3) analogs of $\expect{\HO_{\intr}}$ at $\beta=0$ are calculated using the
appropriate CC energy formulas [Eq. (\ref{energytwobody}) for CCSD and
Eq. (\ref{crcc23}) for CR-CC(2,3)] applied to $H = \HO_{\intr}$, rather than the expectation values
of $\HO_{\intr}$ with the CC wave functions which, as explained in Section \ref{sec:cc},
are not used in the practical implementations of CC methods employed in this work.
The CCSD and CR-CC(2,3) $\expect{\HO_{\cm}}$ values are obtained by numerically differentiating the
corresponding energies, Eq. (\ref{energytwobody}) for CCSD and Eq. (\ref{crcc23}) for CR-CC(2,3),
where $H = H_{\beta}$, with respect to $\beta$ at the $\beta$ values of interest, as in the response
CC considerations described in the previous section. Following the Lawson recipe, the CCSD and CR-CC(2,3) values of
$\expect{\HO_{\intr}}$ at nonzero $\beta$ are calculated as $E_{0}(\beta) - \beta \expect{\HO_{\cm}}_{\beta}$,
where $E_{0}(\beta)$ is the relevant CC [CCSD or CR-CC(2,3)] energy obtained for the
CM-corrected Hamiltonian $H_{\beta}$ and $\expect{\HO_{\cm}}_{\beta} = \partial E_{0}(\beta)/\partial \beta$
is the corresponding value of $\expect{\HO_{\cm}}$ calculated at the same $\beta$ through
the numerical differentiation of $E_{0}(\beta)$, as described above.

\begin{figure}
\includegraphics[width=0.85\columnwidth]{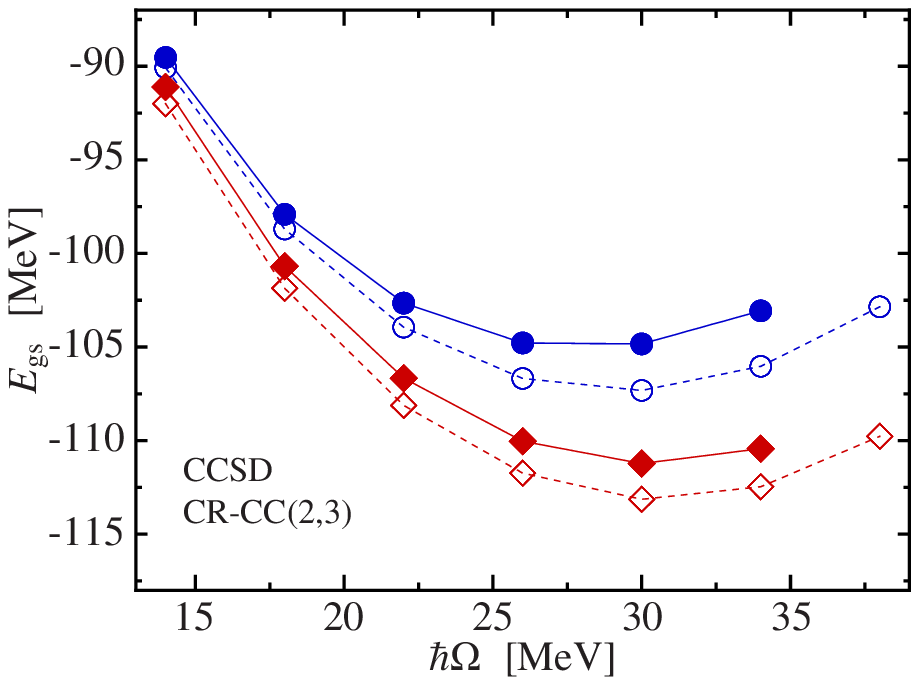}
\caption{(color online) The intrinsic ground-state energies, $\expect{\HO_{\intr}}$,
of \elem{O}{16}, obtained with the CCSD and CR-CC(2,3) approaches at $\beta=0$ (open symbols) and
$\beta=10$ (filled symbols), as functions of $\hbar\Omega$ for a harmonic oscillator basis
with $e_{\max}=5$: CCSD (\symbolcircle[FGBlue],\symbolcircleopen[FGBlue])
and CR-CC(2,3) (\symboldiamond[FGRed],\symboldiamondopen[FGRed]).}
\label{fig:lamHcm}
\end{figure}

As Table \ref{tab:cm_comparison} shows, the full NCSM approach allows an exact separation of intrinsic and CM motions and consequently the intrinsic energy $\expect{\HO_{\intr}}$ is independent of $\beta$. The IT-NCSM(4p4h) method shows a minimal variation of the intrinsic energy at the level of $0.1\MeV$, which is a consequence of the importance truncation. This tiny coupling in the IT-NCSM(4p4h) calculations can safely be neglected. The CCSD, CR-CC(2,3), and IT-CI(4p4h) methods violate translational invariance from the beginning through the choice of the model space.
This results in a more sizable coupling between intrinsic
and CM motions when compared to the IT-NCSM(4p4h) scheme:
the intrinsic energies $\expect{\HO_{\intr}}$ change by $2$ to $3 \MeV$ when going from $\beta=0$ to $\beta=10$.
As mentioned above,
the oscillator frequencies used in Table \ref{tab:cm_comparison} correspond to the approximate
positions of the minima
on the curves that show the dependencies of the relevant energies on $\hbar\Omega$.
The full $\hbar\Omega$-dependence of the intrinsic energies $\expect{\HO_{\intr}}$ for
$\beta=0$ and $\beta=10$ obtained in the CCSD and CR-CC(2,3) calculations is
depicted in Fig. \ref{fig:lamHcm}. Evidently, the change of $\expect{\HO_{\intr}}$,
when going from $\beta=0$ to $\beta=10$, increases somewhat with increasing oscillator frequency.
For example,
for $\hbar\Omega = 18\MeV$, the intrinsic energy obtained with the CR-CC(2,3) approach changes by 1.18 MeV
when going from $\beta=0$ to $\beta=10$, whereas the analogous change in the CR-CC(2,3) energy
for $\hbar\Omega = 34\MeV$ is 2.03 MeV.

The above results demonstrate that the Lawson prescription is quite efficient in reducing the expectation value of the
CM Hamiltonian, $\expect{\HO_{\cm}}$, obtained from the CCSD, CR-CC(2,3), and IT-CI(4p4h) solutions,
as shown in Table \ref{tab:cm_comparison}.
However, this does not mean that a decoupling between intrinsic and CM motions or a
spuriousness-free intrinsic state has been completely achieved. In fact, if we take the same
Lawson prescription and compare the intrinsic energies $\expect{\HO_{\intr}}$ for $\beta=10$ and $\beta=20$,
we observe changes in the values of $\expect{\HO_{\intr}}$
which are similar in magnitude to those observed when one goes from $\beta=0$ to $\beta=10$.
For example, for the IT-CI(4p4h) calculations at $\hbar\Omega = 30$ MeV using 6 major oscillator shells (i.e. $e_{\max}=5$),
the change in the $\expect{\HO_{\intr}}$ value when going from
$\beta=0$ to $\beta=10$ is 1.35 MeV. When going from $\beta=10$ to $\beta=20$, the
IT-CI(4p4h) value of the $\expect{\HO_{\intr}}$
energy changes by 0.91 MeV although $\expect{\HO_{\cm}}$ at $\beta=10$ is already below 0.2 MeV.
This shows that the smallness of $\expect{\HO_{\cm}}$ alone does not indicate a decoupling
or warrant an intrinsic state free of CM contaminations. 

Apart from an explicit projection \cite{RaFa90} or a translationally invariant formulation,
there is no rigorous way to eliminate the CM contamination problem from the CC and CI calculations.
Out of the methods presented in this work, only the NCSM and IT-NCSM calculations are free
(in the IT-NCSM case, virtually free)
of this spuriousness. The CC and CI results are CM contaminated although, as shown in Table \ref{tab:cm_comparison}
and Fig. \ref{fig:lamHcm},
the degree of this contamination, when the harmonic oscillator reference $|\Phi_{0}\rangle$ is used,
seems relatively small in the case of ${^{16}}$O. Indeed, the
degree of CM contamination, as measured by the changes in the intrinsic energies
$\expect{\HO_{\intr}}$ when going from $\beta=0$ to $\beta=10$, does not seem to exceed
2--3 MeV when a medium-size single-particle basis consisting of 6 major oscillator shells is employed
and $|\Phi_{0}\rangle$ is the harmonic oscillator reference,
although we must remember that similar changes in the intrinsic energies
are observed when going from $\beta=10$ to $\beta=20$, suggesting that the real degree
of CM contamination in the CC and CI results is somewhat bigger than 2--3 MeV.
The degree of CM contamination, as measured by the $\expect{\HO_{\cm}}$ values at $\beta=0$, does not
seem to exceed about 5--6 MeV in the same basis.
When the size of the single-particle basis is increased, as is done in the following sections where we examine
single-particle basis sets as large as $e_{\max}=7$ (8 major oscillator shells), the magnitude of the
unphysical coupling of intrinsic and CM motions in the CC and CI calculations is expected to
be reduced, since we approach the limit of the complete single-particle basis.
On the other hand, other effects, such as the $\hbar\Omega$-dependencies of the resulting
energies shown in Fig. \ref{fig:lamHcm} and the fact that the CCSD, CR-CC(2,3), and IT-CI(4p4h) methods do
not provide the exact wave function that would factorize into the intrinsic and translational components,
might hinder the reduction of the CM contaminations present in the
CC and CI results. For all these various reasons, in the assessment of the quality of
the CC and CI calculations reported in this work, we will remain cautious and keep in mind that the resulting
ground-state energies may carry an uncertainty anywhere between 2 and 6 MeV or so as a result of CM contamination,
at least when the harmonic oscillator reference $|\Phi_{0}\rangle$ is employed.
We will continue examining the role of CM contaminations on the CCSD, CR-CC(2,3), and IT-CI(4p4h)
calculations with different types of single-particle bases and different basis set sizes in the future work.

\subsection{Role of the single-particle basis}
\label{section4b}

In many cases, the harmonic oscillator basis is not the optimal choice for the expansion of the nuclear many-body state. The naive reference determinant $|\Phi_{0}\rangle$ obtained by occupying the lowest-energy harmonic oscillator states may have a relatively small overlap with the final wave function $|\Psi_{0}\rangle$, resulting in unnecessarily long CI expansions to represent the correlated $|\Psi_{0}\rangle$ state that have to compensate for the deficiencies of the reference determinant $|\Phi_{0}\rangle$. By switching to a single-particle basis optimized for the nucleus under consideration, generated, for example, by a Hartree-Fock calculation, the convergence with respect to the many-particle basis can be significantly enhanced. Generally, this option is not used in the NCSM calculations, since the use of a single-particle basis set other than the harmonic oscillator basis would destroy the mathematical decoupling of intrinsic and CM motions that the NCSM model space guarantees for any finite basis set. However, for the truncated CC or CI approaches, where we do not have this property anyway, we may benefit from the use of optimized single-particle bases.

In order to demonstrate the effect of an optimization of the single-particle basis, we compare
the results of the CC and CI calculations using the harmonic oscillator (HO) and Hartree-Fock (HF) bases.
The latter is obtained from a self-consistent Hartree-Fock calculation using the same intrinsic
Hamiltonian $\HO_{\intr}$ and the same single-particle space as those employed in the subsequent many-body
calculations \cite{RoPa06}. Therefore, the HF optimization can be viewed as a unitary transformation
within the set of single-particle states employed in the many-body calculations. For a full CI calculation
at given $e_{\max}$, where all determinants resulting from a given single-particle basis set are included,
this transformation would not affect the results. The situation changes when one uses the truncated CI and CC
wave function expansions, where the ${\rm HO \rightarrow HF}$ transformation of single-particle states may
have an effect on the resulting energies. As elaborated on below, this effect is expected to be small
in the case of CC calculations, which rely on the exponential form of the wave function that makes the
results of truncated CC calculations virtually invariant with respect to orbital rotations through the
presence of the $\exp(T_{1})$ component in the CC wave operator, as in the Thouless theorem \cite{thouless},
but can be quite significant when the truncated CI expansions are employed, since the
CI wave operator is a linear rather than an exponential excitation operator which does not contain sufficiently
many terms to make the results numerically invariant with respect to orbital rotations if truncated
at the $m$p$m$h excitations with $m \ll A$. Our numerical analysis confirms these expectations.

\begin{figure}
\includegraphics[width=0.85\columnwidth]{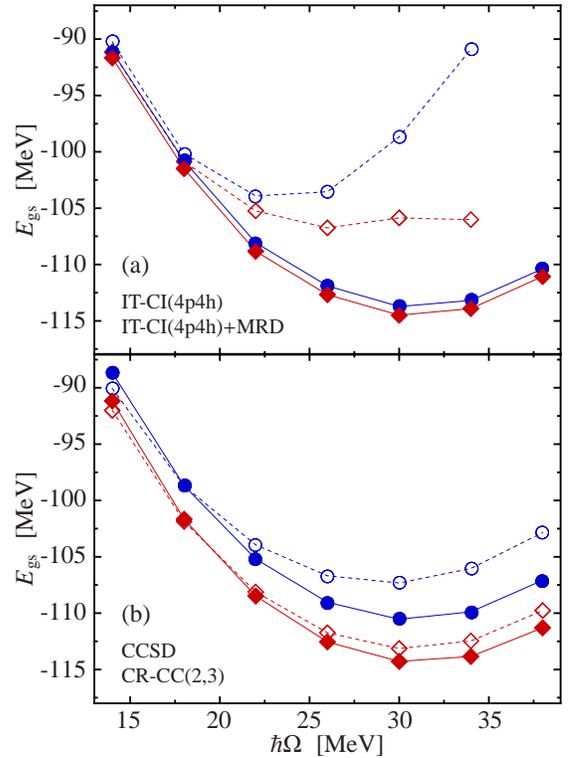}
\caption{(color online) Ground-state energy of \elem{O}{16} as a function of $\hbar\Omega$ obtained with the HO single-particle basis (open symbols) and the HF-optimized basis (filled symbols) in an $e_{\max}=5$ model space. (a) Importance-truncated configuration interaction calculations: IT-CI(4p4h) (\symbolcircle[FGBlue],\symbolcircleopen[FGBlue]) and  IT-CI(4p4h)+MRD (\symboldiamond[FGRed],\symboldiamondopen[FGRed]). 
(b) Coupled-cluster calculations: CCSD (\symbolcircle[FGBlue],\symbolcircleopen[FGBlue]) and CR-CC(2,3) (\symboldiamond[FGRed],\symboldiamondopen[FGRed]).}
\label{fig:spbasis}
\end{figure}

The impact of the basis optimization on a IT-CI(4p4h) calculation is illustrated in
Fig.~\ref{fig:spbasis}(a), where we compare the HO- and HF-based IT-CI(4p4h) results obtained with
the $e_{\max}=5$ model space. We performed similar calculations using other model spaces up to
and including $e_{\max}=7$, and the results are very similar to those shown in Fig.~\ref{fig:spbasis}(a), so
in the following discussion we focus on the $e_{\max}=5$ case.
At smaller oscillator frequencies, the ground-state
energies of $^{16}$O obtained with both bases agree very well.
However, with increasing $\hbar\Omega$ the HF basis leads to lower ground-state energies
than the HO basis. At the same time, the Davidson correction for the HO-based calculation
increases rapidly, indicating that contributions beyond the 4p4h level become significant
in this case.
In contrast to the HO-based IT-CI(4p4h) calculation,
the HF-based IT-CI(4p4h) calculation develops a minimum at larger
$\hbar\Omega$ and the Davidson correction remains small at all frequencies, clearly implying that
the role of higher-than-4p4h excitations in the CI expansion of the ground-state wave function
of $^{16}$O is suppressed when the optimum HF determinant is used as a reference
determinant $|\Phi_{0}\rangle$.

The analogous analysis for the CCSD and CR-CC(2,3) calculations, presented in Fig.~\ref{fig:spbasis}(b), reveals that the CC methods are significantly less sensitive to the choice of the single-particle basis. Again, we only show the sample of the HO- and HF-based CCSD and CR-CC(2,3) calculations corresponding to the $e_{\max}=5$ model space. We performed similar CC calculations using other model spaces up to and including $e_{\max}=7$, and the observed patterns are similar to those shown in Fig.~\ref{fig:spbasis}(b), so we focus on the $e_{\max}=5$ case here. As was the case for the IT-CI(4p4h) calculations, the energies obtained with the two bases are virtually the same for smaller $\hbar\Omega$ values. As we increase the oscillator frequency, the disagreement between the HO- and HF-based CC results grows, but only slightly, particularly for the CR-CC(2,3) approach. Indeed,
unlike the IT-CI(4p4h) case, where the largest discrepancies between the results obtained with the two bases
in the $\hbar\Omega = 14-34$ MeV region, which occur at
$\hbar\Omega = 34\MeV$, are as much as 8 MeV with the Davidson correction and about 20 MeV without it,
the analogous differences between the HO- and HF-based CC results at larger $\hbar\Omega$ values are quite small, with the two results differing by 3.8(4.2) MeV in the CCSD case and 1.4(1.5) MeV in the CR-CC(2,3) case, when the $e_{\rm max} = 5$ basis set is employed and $\hbar\Omega = 34(38)\MeV$. Furthermore, we see that the overall shapes of the curves displaying the $\hbar\Omega$ dependence of the total energies obtained with the CCSD and CR-CC(2,3) approaches are very similar regardless of which single-particle basis is used. It is interesting to note that for both the CC and CI methodologies, the agreement between the results obtained with the two bases improves as we use more accurate approximations. This is particularly true for the CC calculations, where the difference between the HO- and HF-based results obtained with the CR-CC(2,3) method, which is a more accurate approximation when compared to CCSD, is smaller than the analogous difference between the HO- and HF-based CCSD results. The differences between the CI results obtained with the two bases are generally larger, but even in this case, the difference between the HO- and HF-based results obtained with the IT-CI(4p4h)+MRD approach, which corrects the IT-CI(4p4h) energies for at least some effects of higher-than-4p4h excitations, are smaller than the analogous difference between the HO- and HF-based IT-CI(4p4h) results. This makes sense, of course, since the closer we get to the full CI limit, the less sensitive the results become with respect to orbital rotations.

The relative insensitivity of the CC results to the choice of the single-particle basis, which has been
known in quantum chemistry for a long time (cf., e.g., Ref. \cite{relaxation}), is
a consequence of the implicit inclusion of the
Thouless theorem \cite{thouless} in the CC calculations through
the $\exp(T_1)$ component of the CC wave operator
$\exp(T)$, even when $T$ is truncated at the two-body level,
as in CCSD, where $T = T_{1} + T_{2}$. The $\exp(T_1)$ component of the
CCSD wave operator $\exp(T_{1}+T_{2}) = \exp(T_1) \exp(T_2)$ ($T_{1}$ and $T_{2}$ are
particle-hole excitation operators and hence they commute),
obtained by solving the coupled system of equations
for the $T_{1}$ and $T_{2}$ clusters, as described in Section \ref{sec:cc},
acting on the reference determinant $|\Phi_{0}\rangle$
effectively optimizes the single-particle basis for the
many-particle state of interest without the need for the explicit introduction of
orbital relaxation,
thus reducing the impact of any inadequacies of the basis on the final results and making
the CCSD energies almost independent of the type of the single-particle basis.
One can simply write the CCSD wave function, Eq. (\ref{rightccsd}),
as $|\Psi_{0}^{\rm (CCSD)} \rangle = \exp(T_{1}+T_{2}) |\Phi_{0}\rangle
= \exp(T_{2}) |\Phi_{0}^{\prime}\rangle$, where
$|\Phi_{0}^{\prime}\rangle = \exp(T_{1}) |\Phi_{0}\rangle$ is a new reference determinant
optimized for the ground-state $|\Psi_{0}^{\rm (CCSD)}\rangle$ through the suitable choice of
$T_{1}$ obtained from the CCSD calculations.
Indeed, if we
analyze the $T_1$ cluster amplitudes resulting from our CCSD calculations, we see that they are
quite large when the HO basis and large $\hbar\Omega$ are employed,
i.e. when the HO reference determinant $|\Phi_{0}\rangle$ is far from the optimum reference.
When the optimized HF basis is employed, the $T_1$ cluster amplitudes resulting from
CCSD calculations are small, independent of $\hbar\Omega$.
These patterns are reflected in the
values of the so-called `$T_{1}$ Diagnostic' \cite{t1diagnostic}, which is defined as
$(\langle \Phi_{0} | T_{1}^{\dagger} T_{1} | \Phi_{0}\rangle/n_{o})^{1/2}$.
In defining the $T_{1}$ Diagnostic, we
divide the connected (i.e. size extensive)
quantity $\langle \Phi_{0} | T_{1}^{\dagger} T_{1} | \Phi_{0}\rangle$,
which represents the magnitude of $T_{1}$ cluster contributions to the wave function,
by $n_{o}$ to make the result independent of the system size.
For the case of the $e_{\max}=5$ model space and the HO basis,
the $T_{1}$ Diagnostic resulting from the CCSD calculations
changes quite dramatically, from 0.10 at $\hbar\Omega = 14 \MeV$ to 0.42 at $\hbar\Omega = 38 \MeV$ and
0.48 at $\hbar\Omega = 42 \MeV$, indicating a steep increase in the values of the $T_{1}$ cluster
amplitudes due to the increasing inadequacy of the HO reference determinant $|\Phi_{0}\rangle$
at larger $\hbar\Omega$ values that those large $T_{1}$ amplitudes compensate for.
For the optimized HF basis, the same diagnostic remains almost constant,
with values that do not exceed 0.04 in the entire $\hbar\Omega = 14-42 \MeV$ region,
confirming the smallness of $T_{1}$ cluster amplitudes independent of $\hbar\Omega$ in
the HF-basis case.
In fact, it is easy to understand why $T_{1}$ is small in the HF basis. When $|\Phi_{0}\rangle$
is the HF reference, the lowest orders of many-body perturbation theory that
the $T_{1}$ cluster component contributes to
are second for the wave function and fourth for the energy. For comparison, $T_{1}$
contributes to the first-order wave function and second-order correction to the energy when
the non-HF references, such as the HO reference determinant, are used.
In the region of larger $\hbar\Omega$ values, the HO reference determinant is so far from the
optimum HF determinant that the resulting $T_{1}$ amplitudes become very large, as reflected in the
above values of the $T_{1}$ Diagnostic.

In order to confirm that the primary role of the $T_{1}$ operator
in CC calculations is to effectively relax
the orbitals to produce the optimum reference determinant
$|\Phi_{0}^{\prime}\rangle = \exp(T_{1}) |\Phi_{0}\rangle$ for the many-particle state of interest,
we compare the $T_{2}$ cluster operators resulting from the HO- and HF-based CCSD calculations.
We expect the $T_{2}$ clusters, which describe the leading correlation effects,
to be very similar in the HO- and HF-based CCSD calculations if
the main role of $T_{1}$ is to optimize the reference determinant.
The `$T_{2}$ Diagnostic' (cf., e.g., Ref. \cite{ni56_2007}), which is defined as
$(\langle \Phi_{0} | T_{2}^{\dagger} T_{2} | \Phi_{0}\rangle/n_{o})^{1/2}$
and which measures the significance of the $T_{2}$ cluster contributions,
confirms this expectation. For the $e_{\max}=5$ model space,
the values of the $T_{2}$ Diagnostic obtained from the
HO-based CCSD calculations are 0.17--0.18 in the entire 
$\hbar\Omega = 14-42 \MeV$ region. The analogous values of the $T_{2}$ Diagnostic
resulting from the HF-based CCSD calculations are 0.15--0.17 in the same region.
Thus, the $T_{2}$ clusters that describe the true correlation effects barely change with
the type of the basis (HO vs. HF) and $\hbar\Omega$.
The $T_{1}$ clusters remain small and do not change
much with $\hbar\Omega$ when the optimized HF basis is employed, while becoming
sizable in the large $\hbar\Omega$ region when the naive reference determinant $|\Phi_{0}\rangle$,
obtained by filling the lowest-energy HO single-particle states, which is a
poor representation of the correlated ground state in that region, is employed.
The CC theory can cope with the inadequacy of the HO basis by using the $\exp(T_{1})$
component of the
CC wave operator with large $T_{1}$ amplitudes to
transform the naive reference determinant $|\Phi_{0}\rangle$ resulting from the use of the HO basis,
as in the Thouless theorem, to the more optimal form which is
adjusted to the ground-state wave function $|\Psi_{0}\rangle$. The same $\exp(T_{1})$ operator does not
change the reference determinant $|\Phi_{0}\rangle$ too much when $T_{1}$ is small, i.e. when
the orbitals are properly optimized beforehand, as in the HF case. This explains the virtual
invariance of the CC results on the choice of the single-particle basis.

The same arguments enable us to understand why the results of truncated CI calculations may significantly depend on the type of the basis in the region of larger $\hbar\Omega$ values and why one needs to use the HF-optimized orbitals in that region to obtain the results of the CC quality with the truncated CI approaches. Let us focus on the IT-CI(4p4h) approach and the related CISDTQ scheme, in which the linear excitation operator $C$ defining the ground-state wave function $| \Psi_{0}\rangle$ through the formula $| \Psi_{0}\rangle = C |\Phi_{0}\rangle$ has the truncated form $C = C_{0} + C_{1} + C_{2} + C_{3} + C_{4}$. Here, $C_{k}$ is the $k$p$k$h excitation operator generating the contributions from the $k$-tuply excited determinants when acting on $|\Phi_{0}\rangle$ ($C_{0}$ generates the reference contribution). When the intermediate normalization condition $\langle \Phi_{0}| \Psi_{0}\rangle = 1$ is imposed on the CI wave function, so that $C_{0}$ becomes a unit operator, the 1p1h component of the CI wave function, $C_{1}|\Phi_{0}\rangle$, is equivalent to the 1p1h component of the CC wave function, $T_{1}|\Phi_{0}\rangle$. In addition to the connected $T_{1}|\Phi_{0}\rangle$ contribution, the IT-CI(4p4h) and CISDTQ wave functions contain the disconnected cluster terms, such as $(1/2)T_{1}^{2}|\Phi_{0}\rangle$ (through the $C_{2}$ contribution), $(1/6)T_{1}^{3}|\Phi_{0}\rangle$ (through the $C_{3}$ contribution), and $(1/24)T_{1}^{4}|\Phi_{0}\rangle$ (through the $C_{4}$ contribution), but they do not contain the entire $\exp(T_{1}) |\Phi_{0}\rangle$ expansion, which includes higher powers of $T_{1}$ if $A > 4$ (as is the case for the ${^{16}}$O nucleus). In other words, the linear excitation operators $C$ of the IT-CI(4p4h) and CISDTQ approaches or other truncated CI schemes do not have the mathematical structure of the Thouless theorem that would enable one to factor out the $\exp(T_{1})$ component which would make the results virtually independent of the orbital choice. In consequence, the results of truncated CI calculations may display a strong dependence on the choice of the basis, as our IT-CI(4p4h) calculations shown in Fig.~\ref{fig:spbasis} clearly demonstrate. Full CI is the only CI method that contains the $\exp(T_{1})|\Phi_{0}\rangle$ component in its entirety, since one can always represent the intermediately normalized full CI wave function for the $A$-body system, $| \Psi_{0}\rangle = (1 + C_{1} + \cdots + C_{A}) |\Phi_{0}\rangle$, in the exponential form $| \Psi_{0}\rangle = \exp(T_{1} + \cdots + T_{A}) |\Phi_{0}\rangle$.

The analogous analysis can be used to explain why the HO-based IT-CI(4p4h) calculations become less accurate in the region of larger $\hbar\Omega$ values and why the IT-CI(4p4h) calculations benefit from the use of the HF-optimized orbitals, particularly in the region of larger $\hbar\Omega$ values. As already pointed out, the $T_{1}$ cluster component becomes large in the region of larger $\hbar\Omega$ values when the HO reference $|\Phi_{0}\rangle$ is employed. This can be seen by analyzing the CCSD wave function, as described above, or by examining the IT-CI(4p4h) wave function. Since the 1p1h component of the CI wave function, $C_{1}|\Phi_{0}\rangle$, is equivalent to the 1p1h component of the CC wave function, $T_{1}|\Phi_{0}\rangle$, when the intermediate normalization is imposed, we can immediately conclude that if $T_{1}$ is large, as is the case for larger $\hbar\Omega$ values and HO basis, the corresponding $C_{1}$ excitation amplitudes should be large as well. This is exactly what we observe in the HO-based IT-CI(4p4h) calculations. For example, the largest $C_{1}$ excitation amplitude in the IT-CI(4p4h) intermediately normalized wave function increases from 0.08 at $\hbar\Omega = 14$ MeV to 0.29 at $\hbar\Omega = 30$ MeV and 0.40 at $\hbar\Omega = 38$ MeV, when the HO basis is used and $e_{\max}=5$. In contrast, the HF-based IT-CI(4p4h) calculations exhibit  largest $C_1$ amplitudes which do not exceed 0.03 throughout the entire $\hbar\Omega$-range. Now, if the $C_{1}$ or $T_{1}$ component is large, the product terms such as $T_{1}T_{2}$, $(1/2)T_{1}^{2}T_{2}$, $(1/6)T_{1}^{3}T_{2}$, $(1/2)T_{1}T_{2}^{2}$, $(1/4)T_{1}^{2}T_{2}^{2}$, etc. become relatively large as well. All of these terms are included in the CCSD wave function, helping the accuracy of the CCSD and CR-CC(2,3) calculations, but not all of them are present in the IT-CI(4p4h) wave function, which contains the $T_{1}T_{2}$ and $(1/2)T_{1}^{2}T_{2}$ components through the 3p3h and 4p4h contributions described by $C_{3}$ and $C_{4}$, respectively, but not the $(1/6)T_{1}^{3}T_{2}$, $(1/2)T_{1}T_{2}^{2}$, and $(1/4)T_{1}^{2}T_{2}^{2}$ components, which represent the 5p5h [$(1/6)T_{1}^{3}T_{2}$ and $(1/2)T_{1}T_{2}^{2}$] and 6p6h [$(1/4)T_{1}^{2}T_{2}^{2}$] excitations neglected in IT-CI(4p4h). As we can see, the absence of the 5p5h, 6p6h, etc. components in the IT-CI(4p4h) wave function hurts the accuracy of the IT-CI(4p4h) calculations when $C_{1}$ or $T_{1}$ is large, which is exactly what happens when we use the HO basis in the region of larger $\hbar\Omega$ values.

The situation dramatically changes when the HF basis is employed. In that case, the $C_{1}$ or $T_{1}$ contributions are small and the higher-order product terms, such as $(1/6)T_{1}^{3}T_{2}$, $(1/2)T_{1}T_{2}^{2}$, and $(1/4)T_{1}^{2}T_{2}^{2}$, which are neglected in the IT-CI(4p4h) calculations, become very small as well, resulting in an excellent description of the ${^{16}}$O nucleus by the IT-CI(4p4h) method in the entire region of $\hbar\Omega$ which matches the accuracy of the CR-CC(2,3) calculations, particularly when the already very good IT-CI(4p4h) results are corrected for size extensivity and the remaining small higher-than-4p4h excitations through the multi-reference Davidson correction. The excellent agreement between the HF-based IT-CI(4p4h)+MRD and CR-CC(2,3) results, illustrated in Fig.~\ref{fig:spbasis} for $e_{\rm max} = 5$, remains valid when larger model spaces are employed, enabling us to draw several important conclusions regarding the quality of the {\it ab initio} results for ${^{16}}$O reported in this work. These conclusions are discussed in the next section.

\subsection{Comparison of large-scale calculations}
\label{section4c}

We now compare the predictions for the ground-state energy
of \elem{O}{16} obtained in the IT-NCSM, IT-CI, and CC calculations
employing $\VO_{\UCOM}$ and the largest model spaces
that we could handle with a reasonable computational effort.

\begin{figure}
\includegraphics[width=0.85\columnwidth]{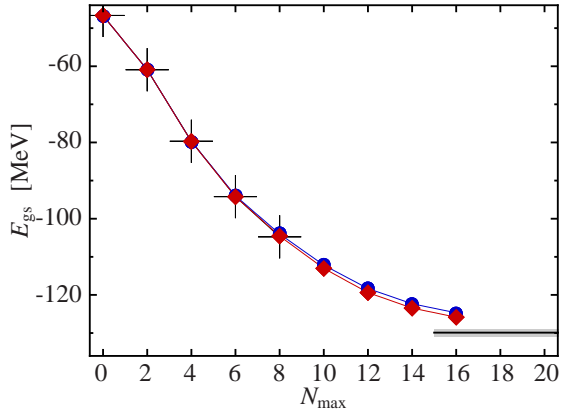}
\caption{(color online) Importance-truncated no-core shell model calculations for the ground-state energy of \elem{O}{16} with $\VO_{\UCOM}$ for $\hbar\Omega=22\MeV$. Shown are the results of the IT-NCSM(4p4h) calculations (\symbolcircle[FGBlue]), IT-NCSM(4p4h)+MRD calculations (\symboldiamond[FGRed]), and full NCSM calculations ($+$). The vertical line indicates the result of an exponential extrapolation of the IT-NCSM(4p4h)+MRD energies.}
\label{fig:convergence_ncsm}
\end{figure}

In Fig. \ref{fig:convergence_ncsm} the convergence of the IT-NCSM ground-state energy as a function of $N_{\max}$ is shown for fixed $\hbar\Omega=22\MeV$, which is determined by the position of the energy minimum for the larger model spaces \cite{Roth08b}. In addition to the IT-NCSM(4p4h) results without and with the Davidson correction, we report the results of full NCSM calculations using the \textsc{Antoine} code \cite{CaNo99} for $N_{\max}\leq8$. In the region where these exact reference results are available, the IT-NCSM(4p4h)+MRD and full NCSM energies agree to within $0.1\MeV$ or better.

\begin{figure}
\includegraphics[width=0.85\columnwidth]{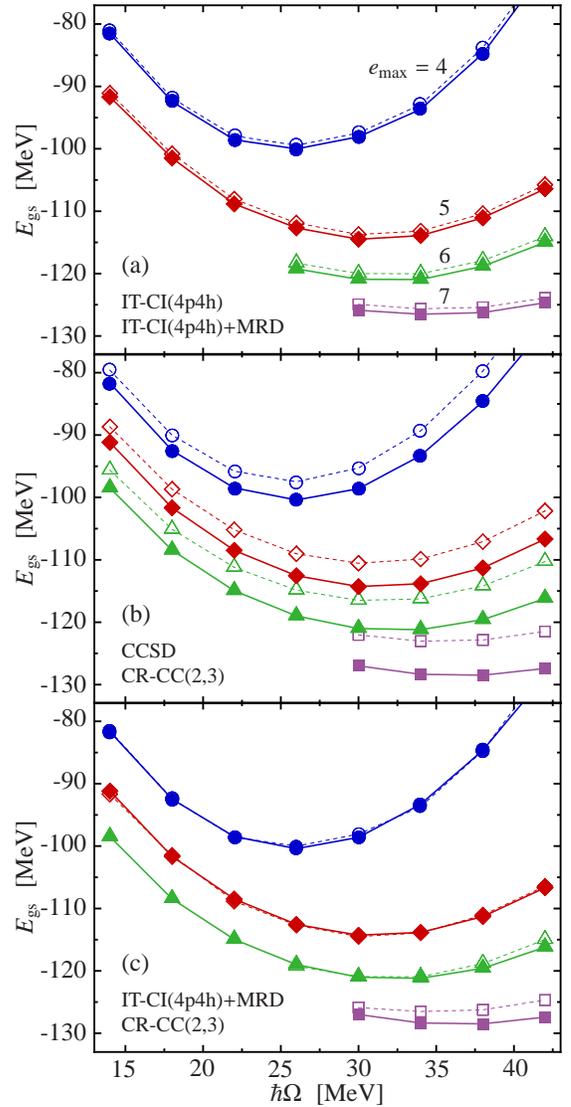}
\caption{(color online) Systematic comparison of IT-CI and CC results for the ground-state energy of \elem{O}{16} using HF-optimized single-particle bases with $e_{\max}=4,5,6$, and $7$. (a) Comparison of IT-CI(4p4h) (open symbols) with IT-CI(4p4h)+MRD (filled symbols). (b) Comparison of CCSD (open symbols) with CR-CC(2,3) (filled symbols). (c) Comparison of IT-CI(4p4h)+MRD (open symbols) with CR-CC(2,3) (filled symbols).}
\label{fig:convergence_cicc}
\end{figure}

Using exponential fits involving the five largest model spaces \cite{reference-note},
we obtain an extrapolated ground-state energy of $(-129\pm1) \: \MeV$ for the IT-NCSM(4p4h) data and of  $(-130\pm1) \: \MeV$ for the IT-NCSM(4p4h)+MRD results. The change due to the Davidson correction provides an estimate of the effects beyond 4p4h configurations. This estimate agrees very well with the preliminary results of the explicit inclusion of up to 6p6h configurations which will be fully elaborated on elsewhere \cite{Roth08b}. The comparison of these results with the experimental binding energy of $-127.6 \MeV$ \cite{AuWa95} proves that the $\VO_{\UCOM}$ two-body interaction provides an excellent description of ground-state energies for heavier nuclei. 

The results of the IT-CI(4p4h) and IT-CI(4p4h)+MRD calculations based on the HF-optimized basis are summarized in Fig. \ref{fig:convergence_cicc}(a). With increasing size of the single-particle basis from $e_{\max}=4$ to $7$ the position of the energy minimum shifts systematically towards larger $\hbar\Omega$. The Davidson correction remains on the order of $1 \MeV$ for all model-space sizes and oscillator frequencies, indicating that the effects of beyond-4p4h configurations are small when the HF basis set is employed. A similar picture emerges from the CC calculations, shown in Fig. \ref{fig:convergence_cicc}(b), which use the same HF-optimized bases and the same model spaces as the IT-CI(4p4h) and IT-CI(4p4h)+MRD calculations. The inclusion of connected triples through the CR-CC(2,3) scheme leads to a lowering of the ground-state energy by up to $6\MeV$, indicating the importance of $T_{3}$ cluster contributions in the quantitative calculations of the nuclear binding energies.

A direct comparison of the IT-CI(4p4h)+MRD and CR-CC(2,3) results
is presented in Fig. \ref{fig:convergence_cicc}(c). The agreement between these
two entirely different many-body approaches is extraordinary. Apart from the largest
model space employed in this study consisting of 8 major oscillator shells,
the two data sets are practically on top of each other.
For the largest $e_{\max}=7$ space, 
the discrepancies between the CR-CC(2,3) and IT-CI(4p4h)+MRD results
are slightly larger than in the case of smaller basis sets,
i.e. the CR-CC(2,3) energies are approximately $1-2\MeV$ lower than the
corresponding IT-CI(4p4h)+MRD energies, but the
overall agreement between the IT-CI(4p4h)+MRD and CR-CC(2,3) energies is outstanding.
Based on this systematic agreement, we can conclude that neither the lack of
strict size extensivity of the truncated IT-CI(4p4h) calculations, which can be
taken care of through the use of the Davidson corrections, nor the violation of the
variational principle by the CC methods, which is compensated by the
high accuracy these methods offer, pose significant practical problems.

The excellent agreement between the CR-CC(2,3) and IT-CI(4p4h)+MRD data can
be rationalized by comparing the CC and CI wave function expansions. If we
impose the intermediate normalization condition $\langle \Phi_{0} | \Psi_{0} \rangle = 1$
exploited in CC theory on the exact CI wave function expansion, we obtain the following
relationships between the CI excitation operators $C_{n}$ and the
CC cluster components $T_{n}$:
\begin{eqnarray}
C_{1} & = & T_{1},
\\
C_{2} & = & T_{2} + \half T_{1}^{2},
\\
C_{3} & = & T_{3} + T_{1}T_{2} + \six T_{1}^{3},
\\
C_{4} & = & T_{4} + T_{1} T_{3} + \half T_{2}^{2} + \half T_{1}^{2} T_{2} + \tfour T_{1}^{4}, \;\; {\rm etc.}
\end{eqnarray}
By design, the CR-CC(2,3) approach
provides a highly accurate description of the connected $T_{1}$, $T_{2}$, and $T_{3}$ clusters,
but not of $T_{4}$ or $T_{n}$ with $n > 4$,
and of all of the disconnected product terms that enter the $C_{n}$ excitation operators with $n=1-4$,
except for $T_{1} T_{3}$. The $T_{1} T_{3}$ term is much smaller
than the leading 4p4h component represented by $(1/2) T_{2}^{2}$, particularly when the HF basis
is employed. When the HF basis is employed, $(1/2) T_{2}^{2}$ contributes to the second-order
many-body perturbation theory correction to the wave function and the fourth-order
correction to the energy, whereas the lowest-order corrections to the wave function and energy
resulting from the $T_{1} T_{3}$ cluster are fourth and sixth, respectively.
The connected $T_{4}$ cluster contributions, which contribute to the fifth and
higher orders in the many-body perturbation theory expansion for the energy, are
much smaller than the disconnected $(1/2) T_{2}^{2}$ contributions as well. In fact,
much of the success of CC theory in areas such as quantum chemistry is related to the
fact that one can safely neglect $T_{4}$ in calculations for the non-degenerate closed-shell systems.
The negligible role of $T_{4}$ clusters has also been observed in the study of
the semi-closed-shell $^{56}$Ni nucleus,
as described by the effective Hamiltonian in the $pf$-shell basis \cite{ni56_2007}.
The $^{16}$O nucleus is a closed-shell system, so one does not need $T_{4}$ to
accurately describe its ground state. This is why the CR-CC(2,3) approach
provides a virtually exact description of the CI excitation contributions $C_{n}$ up to and including
the 4p4h (i.e. $n=4$) terms. Since the CR-CC(2,3) method is based on the idea of correcting the
CCSD energy for the leading $T_{3}$ contributions and since the CCSD approach describes
all higher-than-4p4h excitations that can be represented as products of
the $T_{1}$ and $T_{2}$ clusters, one has to correct the IT-CI(4p4h) energies for the
selected higher-than-4p4h correlations via the Davidson corrections to improve the agreement
between the IT-CI(4p4h) and CR-CC(2,3) data. This explains why the IT-CI(4p4h)+MRD and
CR-CC(2,3) results obtained in this work agree so well.

In order to compare the IT-CI(4p4h)+MRD and CR-CC(2,3) results with the aforementioned
converged IT-NCSM calculations, an extrapolation $e_{\max}\to\infty$ is necessary.
In view of the convergence pattern of the CI and CC results and the fact that our
IT-CI(4p4h)+MRD and CR-CC(2,3) data are limited to $e_{\max} \leq 7$,
this extrapolation can only provide a rough estimate. The crude
exponential extrapolations based on the total energies obtained with
the four different model space sizes with $e_{\max}=4-7$
at fixed $\hbar\Omega=30,34,38,$ and $42$MeV, for which the full set of the
IT-CI(4p4h)+MRD and CR-CC(2,3) data for $e_{\max}=4-7$ is available,
lead to an estimate of the $e_{\max}\to\infty$ CR-CC(2,3) energy
in the range of $-131$ to $-133 \: \MeV$ in the entire
$\hbar\Omega = 30-42$ MeV region. The IT-CI(4p4h)+MRD result is similar.
It looks as though the $e_{\max}\to\infty$ CR-CC(2,3) energies
of $-131$ to $-133\MeV$ and their IT-CI(4p4h)+MRD analogs, obtained via the above
exponential extrapolations based on the total energies,
are in excellent agreement with the extrapolated IT-NCSM(4p4h)+MRD result of
$(-130\pm1) \: \MeV$, but we should be very careful in interpreting this agreement,
which might be fortuitous, since crude extrapolations
based on the limited set of CC and CI data that we have access to may carry sizable error bars.
For example, if we use the more careful approach where instead of
extrapolating total energies we extrapolate only the correlation energies obtained with $e_{\rm max} = 3$,
5, and 7, and add the results to the highly accurate estimate of the converged (to within 0.1 MeV) HF energy
resulting from the $e_{\rm max} = 20$ HF calculations, we obtain the extrapolated total CR-CC(2,3) energy
in the range of $-135$ to $-141$ MeV in the entire $\hbar\Omega = 30-42$ MeV region. The analogous
IT-CI(4p4h)+MRD energies are in the range of $-132$ to $-135$ MeV.
The use of only odd values of $e_{\rm max}$ in the correlation energy extrapolations can be justified
by the fact that the HF and, in consequence, correlation energies do not change uniformly
when increasing the basis set; changes in the HF energies are much stronger when another
radial excitation is added to the p states, as observed when going from $e_{\rm max}=4$ to 5
and from $e_{\rm max}=6$ to 7.
Thus, based on the limited set of the CC and CI data we have at our
disposal, we can only state that the extrapolated $e_{\rm max}
\rightarrow \infty$ CR-CC(2,3) and IT-CI(4p4h)+MRD energies fall within
the broader range of $-131$ to $-141$ MeV, which implies that our
extrapolated results carry an uncertainty which could be as big as about
10 MeV.
Clearly,
the few MeV differences between the extrapolated CR-CC(2,3)/IT-CI(4p4h)+MRD and
IT-NCSM(4p4h)+MRD results for the binding
energy of \elem{O}{16} prompt further study, so that we can understand the nature of these differences
in more precise terms, but it is already worth pointing out that these few MeV differences between
the extrapolated CR-CC(2,3) and IT-CI(4p4h)+MRD binding energies on the one hand and their $N_{\max}\to\infty$
IT-NCSM(4p4h)+MRD counterpart on the other hand are consistent with our estimate of the effect of CM
contaminations on the calculated CR-CC(2,3) and IT-CI(4p4h) energies discussed in Section \ref{section4a}.
Since the CR-CC(2,3), IT-CI(4p4h)+MRD, and IT-NCSM(4p4h)+MRD
methods contain similar many-body correlation effects as $e_{\rm max}(N_{\rm max}) \rightarrow \infty$
and since the IT-NCSM(4p4h)+MRD results are virtually free of the CM contaminations
(cf. Section \ref{section4a}), it is quite possible that the CM contaminations in the CR-CC(2,3)
and IT-CI(4p4h)+MRD results are largely responsible for the observed few MeV differences between
the extrapolated CR-CC(2,3) or IT-CI(4p4h)+MRD and IT-NCSM(4p4h)+MRD energies.
The fact that the relatively inexpensive CR-CC(2,3)
approach can produce the binding energy of $^{16}$O which differs from
the best IT-NCSM(4p4h)+MRD estimate and experiment by only a few MeV (less than 10 \%), when
the $\VO_{\UCOM}$ interaction is employed,
is an indication that the
CR-CC(2,3) method captures practically all correlations relevant for the
description of the ground state of $^{16}$O and that $\VO_{\UCOM}$ provides
the accurate representation of the relevant nucleon-nucleon interactions.

\section{Conclusions}

Through the direct comparison of results for the \elem{O}{16} ground-state
energy obtained using the $V_{\UCOM}$ interaction within the IT-NCSM, IT-CI,
and CC approaches, we have established a comprehensive picture of the quality
of the different many-body approaches and the practical relevance of formal
limitations associated with each one of them. Among the points that we have discussed
in detail are the possible role of CM contaminations, the choice of the
single-particle basis, and the impact of size-extensivity.

The analysis of the coupling of intrinsic and CM motions using the Lawson prescription shows that
the IT-CI and CC methods, which are based on a single-particle truncation when constructing
the relevant model spaces, exhibit a coupling
between intrinsic and CM motions that in the case of \elem{O}{16} affects the intrinsic energies
at the level of about 2--6 MeV or so when the HO reference determinant is employed.
The same analysis also indicates that a small value of the
expectation value of the CM Hamiltonian
$\expect{\HO_{\cm}}$ alone does not automatically warrant a decoupling and a spuriousness-free intrinsic state.
Only the IT-NCSM approach, which is based on approximating the complete $N_{\max}\hbar\Omega$ model space of
the NCSM theory, shows a virtually perfect decoupling, leading
to effectively contamination-free intrinsic eigenstates.
On the other hand, the relatively small CM contaminations observed in the IT-CI and CC calculations for
\elem{O}{16} do not seem to be detrimental for the quality of the resulting energies, which almost
perfectly agree with one another when
the appropriate levels of both theories are employed, namely, IT-CI(4p4h)+MRD in the case of IT-CI and
CR-CC(2,3) in the case of CC, and which are in reasonable
agreement with the results of the converged IT-NCSM(4p4h)+MRD calculations when we attempt
to extrapolate the IT-CI(4p4h)+MRD and CR-CC(2,3) results
to the complete basis set limit. The remaining few (up to about 10) MeV differences between the crudely extrapolated
CR-CC(2,3) and IT-CI(4p4h)+MRD energies and the converged IT-NCSM(4p4h)+MRD calculations
are consistent with the magnitude of the CM contaminations present in the
CR-CC(2,3) and IT-CI(4p4h)+MRD calculations.

The comparison of the calculations employing the HO single-particle bases with the calculations using
HF-optimized bases demonstrates that the CC method is largely insensitive to the basis set choice,
whereas the IT-CI calculations show a strong basis set dependence, significantly benefiting from the use of
an optimized basis. This fundamentally different behavior of the CC and IT-CI approaches with regard to
the choice of the single-particle basis is related to the presence of
the $\exp(T_{1})$ component in the exponential wave operator of CC theory, as in the Thouless theorem,
which makes the CC results almost insensitive to the basis set choice, and the incomplete
treatment of this component by the linear wave operator of IT-CI.
As shown in this work,
an effective measure of the suitability of the
single-particle basis for the truncated CI (e.g. IT-CI) calculations can be provided by the $T_1$ or $C_1$
excitation amplitudes and the corresponding $T_{1}$ Diagnostic
as well as the magnitude of the Davidson extensivity corrections. The $T_1$ or $C_1$ values,
the values of the $T_{1}$ Diagnostic, and the Davidson extensivity corrections all become
large if the basis is ill-adapted to the truncated CI calculations of interest.
We can, therefore, check the suitability of a given basis set for the
IT-CI and IT-NCSM calculations by monitoring these quantities. 

When using the HF-optimized basis in large scale CR-CC(2,3) and IT-CI(4p4h)+MRD calculations,
we observe an excellent agreement of the ground-state energies
of \elem{O}{16} for all values of $\hbar\Omega$ and $e_{\max}$.
Only for the largest $e_{\max}=7$ space and large $\hbar\Omega$ values
do we observe the slightly larger differences between the CR-CC(2,3) and IT-CI(4p4h)+MRD results,
on the order of 1 to 2 MeV. This excellent agreement between the
results of the CR-CC(2,3) and IT-CI(4p4h)+MRD calculations
demonstrates that neither the violation of strict size extensivity by the IT-CI schemes
nor the violation of the variational principle by the truncated CC schemes are of major practical concern
in nuclear structure calculations, since both the IT-CI and CC methodologies are systematically
improvable through the inclusion of higher-order many-body components in the corresponding excitation operators
($C_{m}$ components in the case of IT-CI and cluster components $T_{m}$ in the case of CC)
and the use of suitable energy corrections (the Davidson corrections in the case of IT-CI and the corrections due
to the effects of higher-order clusters in CC).
The IT-CI(4p4h)+MRD and the CR-CC(2,3) results converge toward somewhat lower binding energies
than the IT-NCSM(4p4h)+MRD calculations, but, as already pointed out, the observed few MeV or
a few percent differences between the
extrapolated $e_{\max} \rightarrow \infty$ IT-CI(4p4h)+MRD and CR-CC(2,3) energies on the one hand and the
$N_{\max} \rightarrow \infty$ IT-NCSM(4p4h)+MRD energies
on the other hand are consistent with the effects expected from the presence of CM
contaminations in the CI and CC calculations. Based on all of these observations, we
conclude that all {\it ab initio} schemes used in the present work --- IT-NCSM, IT-CI, and CC --- provide
powerful, affordable, and potentially accurate computational tools to tackle the nuclear
many-body problem. Due to the complementarity of the IT-NCSM, IT-CI, and CC methods,
comparative computational studies using all of these approaches,
following the analysis presented in this work,
may provide a comprehensive and precise picture of nuclear structure of p-shell nuclei and beyond.

Finally, it is interesting to compare the results for the ground-state energy of \elem{O}{16}
obtained in the present study with the $V_{\UCOM}$ two-body interaction
with the earlier recent CC calculations using other two-body interactions.
The extrapolated ground-state energy obtained in the IT-NCSM(4p4h)+MRD
calculations with $V_{\UCOM}$ is $(-130 \pm 1)$ MeV, i.e. within about 2 MeV from the
experimental value of $-127.6$ MeV. Although the CR-CC(2,3) results reported
in this work and obtained with the
same potential are not as well converged with the single-particle basis set,
the attempt to extrapolate the CR-CC(2,3) energies to the $e_{\rm max} \rightarrow \infty$ limit
produced the result which implies overbinding, compared to experiment, on the order of 3--13 MeV.
We believe that most of this overbinding and the analogous overbinding
observed in IT-CI(4p4h)+MRD calculations is due to CM contaminations that affect the CCSD and the
subsequent CR-CC(2,3) calculations as well as the IT-CI(4p4h)+MRD calculations, although future,
more detailed, studies will be required to verify this statement.
The CR-CCSD(T) study reported in Ref. \cite{o16prl}
using the Idaho-A two-body potential \cite{entem2002} produced, after correcting the result for the effect
of the Coulomb interaction, an extrapolated ground-state energy of about $-109.3$ MeV.
A similar result (approximately, $-112$ MeV) was obtained with the two-body component
of the chiral N$^3$LO potential \cite{EnMa03}, which includes the Coulomb interaction directly \cite{o16prl}.
Thus, with these chiral two-body interactions, the CR-CCSD(T) approach underbinds $^{16}$O
by approximately 17 MeV, which was concluded to be due to the effect of missing three-body forces \cite{o16prl}.
Similar underbinding has also been observed in the CC calculations for ${}^{16}$O employing
the Argonne V18 potential reported in Refs. \cite{bogdan1,bogdan2,bogdan4}, where the explicit
inclusion of three-nucleon interactions was used to improve the
agreement with experiment \cite{bogdan1,bogdan2}.
A recent CCSD(T) study based on the two-body $V_{{\rm low} k}$ interaction presented in Ref. \cite{HaDe07}
(CCSD(T) is another non-iterative treatment of $T_{3}$ clusters in CC theory \cite{ragha},
which is less robust and generally less
accurate than the CR-CC(2,3) approach used in the present study
\cite{crccl_jcp,crccl_cpl,crccl_open,crcc23a,crcc23d,crcc23e,crcc23f,crcc23h}),
has led to an extrapolated ground-state energy of $-148.2$ MeV, i.e. the binding energy overestimated by
more than 20 MeV.
These comparisons demonstrate that the $V_{\UCOM}$ interaction, in contrast to many other
two-nucleon interactions, requires no or minimal contributions from a three-body force
in order to provide a reasonably accurate description of the ground-state energy of \elem{O}{16}.
An extension of these investigations to open-shell systems and excited states, including other
p-shell nuclei, will be presented elsewhere.

\section*{Acknowledgments}

This work was supported by the Deutsche Forschungsgemeinschaft through contract SFB 634, the Helmholtz International Center for FAIR within the framework of the LOEWE program launched by the State of Hesse (R.R.), the Chemical Sciences, Geosciences
and Biosciences Division, Office of Basic Energy Sciences, Office
of Science, U.S. Department of Energy (Grant No. DE-FG02-01ER15228; P.P)
and the National Science Foundation's Graduate Research Fellowship (J.R.G).
The coupled-cluster calculations were performed
on the computer systems provided by the
High Performance Computing Center and Department of Chemistry at Michigan State
University.



\begin{thebibliography}{77}
\bibitem{reference-note}
From the set of five results we extract all possible subsets of three data points
and perform the extrapolations. The average
of the extrapolation results is quoted as energy and the standard deviation as error.
\end{thebibliography}
\end{document}